# ALBEDOS, EQUILIBRIUM TEMPERATURES, AND SURFACE TEMPERATURES OF HABITABLE PLANETS

Anthony D. Del Genio[1], Nancy Y. Kiang[1], Michael J. Way[1], David S. Amundsen[1,2], Linda E. Sohl[1,3], Yuka Fujii[4], Mark Chandler[1,3], Igor Aleinov[1,3], Christopher M. Colose[5], Scott D. Guzewich[6], and Maxwell Kelley[1,7]

Short title: Habitable planet albedos and temperatures

[1]NASA Goddard Institute for Space Studies, 2880 Broadway, New York, NY 10025

[2]Department of Applied Physics and Applied Mathematics, Columbia University, New York, NY 10027

[3]Center for Climate Systems Research, Columbia University, New York, NY 10027

[4]Earth-Life Science Institute, Tokyo Institute of Technology, Ookayama, Meguro, Tokyo 152-8550, Japan

[5] NASA Postdoctoral Program, Goddard Institute for Space Studies, New York, NY 10025

[6]NASA Goddard Space Flight Center, Greenbelt, MD 20771

[7]SciSpace LLC, 2880 Broadway, New York, NY 10025

Corresponding author:

Anthony D. Del Genio
NASA Goddard Institute for Space Studies
2880 Broadway
New York, NY 10025
Phone: 212-678-5588
Email: anthony.d.delgenio@nasa.gov

## ABSTRACT

The potential habitability of known exoplanets is often categorized by a nominal equilibrium temperature assuming a Bond albedo of either ~0.3, similar to Earth, or 0. As an indicator of habitability, this leaves much to be desired, because albedos of other planets can be very different, and because surface temperature exceeds equilibrium temperature due to the atmospheric greenhouse effect. We use an ensemble of 3-dimensional general circulation model simulations to show that for a range of habitable planets, much of the variability of Bond albedo, equilibrium temperature, and even surface temperature can be predicted with useful accuracy from incident stellar flux and stellar temperature, two known external parameters for every confirmed exoplanet. Earth's Bond albedo is near the minimum possible for habitable planets orbiting G stars, because of increasing contributions from clouds and sea ice/snow at higher and lower instellations, respectively. For habitable M star planets, Bond albedo is usually lower than Earth's because of near-IR $H_2O$ absorption, except at high instellation where clouds are important. We apply the relationships derived from this behavior to several known exoplanets to derive zeroth-order estimates of their potential habitability. More expansive multivariate statistical models that include currently non-observable parameters suggest that greenhouse gas variations could account for significant variance in albedo and surface temperature, while orbital and surface features can have significant but not clearly consistent effects. We discuss how emerging information from global climate models might resolve some degeneracies and help focus scarce observing resources on the most promising planets.

# 1. INTRODUCTION

The proliferation of newly discovered rocky planets has expanded the focus of exoplanet science from detection to characterization of their potential habitability. Initial assessments of habitability are based on the planet's location relative to some definition of the habitable zone (see Kane et al., 2016). Often, this is reported as a nominal planetary equilibrium temperature $T^n_{eq}$, assuming a reference Bond albedo $A$ ~ 0.3, like Earth, or $A = 0$, an unrealizable value (e.g., Borucki et al., 2012; Anglada-Escudé et al., 2016; Dittmann et al., 2017; Gillon et al., 2017).

A habitable planet's albedo, however, will be different when placed in a different orbit than Earth's and/or in orbit around a different star, even if the planet is very Earth-like, precisely because such planets have water. An important feature of Earth's climate is its existence near the triple point of $H_2O$. Surface ice, surface liquid, and liquid or ice suspended as cloud particles each make non-negligible contributions to Earth's Bond albedo. Water vapor does not, but only because the Sun is a G star that emits primarily at wavelengths at which $H_2O$ does not absorb. Thus, differences in the occurrence of the three phases of water and their interaction with incident starlight, if predictable, can shed light on actual equilibrium temperatures. Furthermore, equilibrium temperature differs from the actual parameter of interest for habitability, the surface temperature, depending on the atmosphere's greenhouse effect, which varies with composition and pressure. Quantifying the relationship between equilibrium and surface temperature for habitable planets is necessary for assessing potential habitability.

Three-dimensional general circulation models (GCMs) have begun to address questions about the conditions necessary for habitability for particular exoplanets (e.g., Turbet et al., 2016, 2018; Boutle et al., 2017; Wolf, 2017, 2018; Del Genio et al., 2018) and a range of hypothetical exoplanets (Leconte et al., 2013; Shields et al., 2013, 2014; Yang et al., 2013, 2014; Wolf and Toon, 2014, 2015; Bolmont et al., 2016; Kopparapu et al., 2016, 2017; Fujii et al., 2017a; Wolf et al., 2017; Haqq-Misra et al., 2018; Jansen et al., 2018; Lewis et al., 2018; Way et al., 2018). Nonetheless, such models are computationally expensive and not generally available to the larger community, and thus only a limited number of simulations have been conducted to date. At the same time, the population of known rocky exoplanets is likely to grow dramatically as a result of upcoming spacecraft missions and new ground-based telescopes, yet the lengthy observing times needed to characterize small planets will limit such efforts to a small fraction of them. How might the meager available information about rocky exoplanets be used to choose the most promising subset for further study given limited observing resources?

This paper describes the first step of an approach to distill the information in GCM simulations into a few simple relationships that can be applied to known or probable rocky planets to refine initial assessments of their potential habitability. It cannot anticipate whether water, or any atmosphere, exists on any given planet whose estimated size and/or mass makes it likely to be rocky. It can only tell us what the plausible range of albedos might be for a given rocky planet if it has water, and whether that range bodes well for its potential habitability.

Section 2 describes the GCM experiments used in our study. Section 3 discusses factors that control Bond albedo, and thus the *actual* equilibrium temperature, for habitable planets and how predictable these may be from sparse information; it also estimates errors in surface temperature given the equilibrium temperature as a predictor. Section 4 demonstrates the use of the predictor by applying it to a number of known exoplanets. Section 5 explores additional parameters that may influence albedo and temperature and discusses how ambiguities that limit our predictive ability might be reconciled with future observations. Section 6 summarizes our findings and suggests future directions.

## 2. GCM ENSEMBLE

We use simulations performed with the ROCKE-3D GCM (Way et al., 2017) in this analysis. All simulations, except one dry "land planet," couple the atmosphere to a dynamic ocean. Ideally one would construct a large ensemble of simulations that systematically vary every relevant external parameter to represent all possible exoplanet climates. Such an ensemble does not yet exist. Instead, we employ a sparse "ensemble of opportunity" of 48 simulations already conducted with ROCKE-3D to illustrate the concept. Most simulations are for "Earth-like" atmospheres, with ~1 bar of $N_2$ as the dominant atmospheric constituent, $CO_2$ and sometimes $CH_4$, and surface water. However we do include several thinner and thicker atmospheres as well as $CO_2$-dominated atmospheres. The full diversity of habitable exoplanets is not represented by the ensemble, but it does include different instellations, stellar types (G and M), planet sizes and gravities, obliquities and spin-orbit states, land-ocean configurations, and ocean properties. In particular, several subensembles produce different albedos for reasons other than instellation, stellar type, or greenhouse gas concentrations. This illustrates degeneracies unrelated to radiative properties that are rarely considered in discussions of exoplanet climates.

Table 1 provides the relevant properties of all the simulated planets. Our ensemble consists of the following general classes of planets:

- *Proxima Centauri b* (Del Genio et al., 2018): 10 simulations using estimated planet size, gravity, stellar spectrum, distance from the star (Anglada-Escudé et al., 2016), and with different surfaces, atmospheric pressures and compositions, instellations, and spin-orbit configurations.

- *GJ 876* (Fujii et al., 2017a): 4 hypothetical aquaplanets in synchronous rotation using Earth size and gravity and incident stellar flux from 0.6-1.2 x Earth's solar constant $S_o$.
- *Hypothetical early Venus* (Way et al., 2016): 3 simulations with Venus' size, gravity, and modern rotation period (243 d), with different surfaces harboring liquid water, and with insolation and spectrum at different points in Venus' past.
- *Early Earth*: Four periods with different insolations, compositions and surfaces; 3 Archean Earths based on Charnay et al. (2013); 1 Huronian snowball Earth; 4 Sturtian equatorial waterbelt Earths (Sohl and Chandler, 2007); 1 mid-Cretaceous equable Earth.
- *Earth rotation-insolation experiments* (Way et al., 2018): 5 simulations with different insolations, rotation periods, and zero obliquity and eccentricity, plus 6 other simulations with Earth's actual obliquity and insolation and different rotation periods.
- *Obliquity experiments* (Colose et al., 2018)*:* 9 Earth aquaplanets with low or high obliquity, and different insolations and greenhouse gas concentrations, including examples with warm and cold start initial conditions that produce bistable behavior.
- *Hypothetical dry early Mars:* Mars size, gravity, and orbital properties with 3.8 Gya instellation and 1 bar $CO_2$, but only trace amounts of water and no $CO_2$ surface ice.
- *Modified Kepler-1649 b analog* (Kane et al., 2018)*:* Aquaplanet in synchronous rotation with estimated planet size and gravity; modern Earth-like atmosphere but weaker instellation and longer orbital period than observed to place it inside the habitable zone.

Most of these planets are habitable based on the criterion of the presence of surface liquid water. The exceptions are Huronian Earth, a complete snowball; the early Mars analog, which has

**Table 1**

Relevant Properties of Simulated Planets[1]

| Simulation | $S_{0X}$ | $T_{star}$ (K) | $p_{surf}$ (b) | Major gases | Trace gases (ppmv) | $P_{rot}$ (d) | Spin-orbit | Surface | Φ (°) | A |
|---|---|---|---|---|---|---|---|---|---|---|
| Proxima Cen b | (Del Genio et al., 2018) | | | | | | | | | |
| (1) Control | 0.65 | 3050 | .984 | $N_2$ | $CO_2$ (376) | 11.2 | 1:1 | Aqua | 0 | .234 |
| (2) Control-Hi | 0.70 | 3050 | .984 | $N_2$ | $CO_2$ (376) | 11.2 | 1:1 | Aqua | 0 | .232 |
| (3) Archean-M | 0.65 | 3050 | .984 | $N_2$ | $CO_2$ (900) $CH_4$ (900) | 11.2 | 1:1 | Aqua | 0 | .181 |
| (4) Archean-H | 0.65 | 3050 | .984 | $N_2$ | $CO_2$ ($10^4$) $CH_4$ (2000) | 11.2 | 1:1 | Aqua | 0 | .161 |
| (5) Hi Salinity | 0.65 | 3050 | .984 | $N_2$ | $CO_2$ (376) | 11.2 | 1:1 | Aqua | 0 | .178 |
| (6) 3:2e30 | 0.65 | 3050 | .984 | $N_2$ | $CO_2$ (376) | 7.5 | 3:2 | Aqua | 0 | .199 |
| (7) Day-Land | 0.65 | 3050 | .984 | $N_2$ | $CO_2$ (376) | 11.2 | 1:1 | Modern Earth | 0 | .253 |
| (8) Cont-Thin | 0.65 | 3050 | 0.1 | $N_2$ | $CO_2$ (376) | 11.2 | 1:1 | Aqua | 0 | .310 |
| (9) Cont-Thick | 0.65 | 3050 | 10 | $N_2$ | $CO_2$ (376) | 11.2 | 1:1 | Aqua | 0 | .237 |
| (10) Pure $CO_2$ | 0.65 | 3050 | 1 | $CO_2$ | None | 11.2 | 1:1 | Aqua | 0 | .222 |
| GJ 876 | (Fujii et al., 2017a) | | | | | | | | | |
| (11) 0.6 | 0.60 | 3129 | 1 | $N_2$ | $CO_2$ (1) | 32.3 | 1:1 | Aqua | 0 | .254 |
| (12) 0.8 | 0.80 | 3129 | 1 | $N_2$ | $CO_2$ (1) | 26.1 | 1:1 | Aqua | 0 | .255 |
| (13) 1.0 | 1.00 | 3129 | 1 | $N_2$ | $CO_2$ (1) | 22.0 | 1:1 | Aqua | 0 | .332 |
| (14) 1.2 | 1.20 | 3129 | 1 | $N_2$ | $CO_2$ (1) | 19.2 | 1:1 | Aqua | 0 | .311 |
| Early Venus | (Way et al., 2016) | | | | | | | | | |
| (15) 1.5 | 1.48 | 5790 | 1.01 | $N_2$ | $CO_2$ (400) $CH_4$ (1) | 243 | Asyn | Venus topo with ocean | 2.6 | .521 |
| (16) 1.9 | 1.90 | 5785 | 1.01 | $N_2$ | $CO_2$ (400) $CH_4$ (1) | 243 | Asyn | Venus topo with ocean | 2.6 | .609 |
| (17) 2.4 | 2.40 | 5785 | 1.01 | $N_2$ | $CO_2$ (400) $CH_4$ (1) | 243 | Asyn | Aqua | 2.6 | .671 |
| Early Earth | (Sohl and Chandler, 2007) | | | | | | | | | |
| (18)Archean A | 0.80 | 5710 | 1.01 | $N_2$ | $CO_2$ (900) $CH_4$ (900) | 1 | Asyn | Modern Earth | 23.5 | .492 |
| (19)Archean B | 0.80 | 5710 | .984 | $N_2$ | $CO_2$ ($10^4$) $CH_4$ (2000) | 1 | Asyn | Modern Earth | 23.5 | .378 |
| (20)Archean C | 0.80 | 5710 | .984 | $N_2$ | $CO_2$ ($10^5$) $CH_4$ (2000) | 1 | Asyn | Sturtian Earth | 23.5 | .286 |
| (21) Huronian | 0.84 | 5728 | .984 | $N_2$ (.79) $O_2$ (.21) | $CO_2$ (40) $CH_4$ (.751) $N_2O$ (.275) | 1 | Asyn | Sturtian Earth | 23.5 | .504 |
| (22) Sturtian 1 | 0.94 | 5760 | .984 | $N_2$ (.79) $O_2$ (.21) | $CO_2$ (40) $CH_4$ (.751) | 1 | Asyn | Sturtian Eartth | 23.5 | .397 |
| | | | | | $N_2O$ (.275) | | | | | |
| (23) Sturtian 2 | 0.94 | 5760 | .984 | $N_2$ (.79)<br>$O_2$ (.21) | $CO_2$ (285)<br>$CH_4$ (.791)<br>$N_2O$ (.275) | 1 | Asyn | Sturtian Earth | 23.5 | .348 |
| (24) Sturtian 3 | 0.94 | 5760 | .984 | $N_2$ (.79)<br>$O_2$ (.21) | $CO_2$ (140)<br>$CH_4$ (.751)<br>$N_2O$ (.275) | 1 | Asyn | Sturtian Earth | 23.5 | .373 |
| (25) Sturtian 4 | 0.90 | 5760 | .984 | $N_2$ (.79)<br>$O_2$ (.21) | $CO_2$ (40)<br>$CH_4$ (.751)<br>$N_2O$ (.275) | 1 | Asyn | Sturtian Earth | 23.5 | .418 |
| (26)Cretaceous | 0.99 | 5773 | .984 | $N_2$ (.79)<br>$O_2$ (.21) | $CO_2$ (40)<br>$CH_4$ (.751)<br>$N_2O$ (.275) | 1 | Asyn | Cretaceous Earth | 23.5 | .295 |
| Earth rot-$S_0$ | (Way et al., 2018) | | | | | | | | | |
| (27) 1/1.1z | 1.1 | 5787 | .984 | $N_2$ | $CO_2$ (400) | 1 | Asyn | Modern Earth | 0 | .348 |
| (28) 1/1.2z | 1.2 | 5787 | .984 | $N_2$ | $CO_2$ (400) | 1 | Asyn | Modern Earth | 0 | .373 |
| (29) 16/1.1z | 1.1 | 5787 | .984 | $N_2$ | $CO_2$ (400) | 16 | Asyn | Modern Earth | 0 | .278 |
| (30) 16/1.2z | 1.2 | 5787 | .984 | $N_2$ | $CO_2$ (400) | 16 | Asyn | Modern Earth | 0 | .314 |
| (31) 16/1.3z | 1.3 | 5787 | .984 | $N_2$ | $CO_2$ (400) | 16 | Asyn | Modern Earth | 0 | .358 |
| (32) 1/1.0 | 1 | 5785 | .984 | $N_2$ (.79)<br>$O_2$ (.21) | $CO_2$ (400)<br>$CH_4$ (1) | 1 | Asyn | Modern Earth | 23.5 | .306 |
| (33) 8/1.0 | 1 | 5785 | .984 | $N_2$ (.79)<br>$O_2$ (.21) | $CO_2$ (400)<br>$CH_4$ (1) | 8 | Asyn | Modern Earth | 23.5 | .294 |
| (34) 16/1.0 | 1 | 5785 | .984 | $N_2$ (.79)<br>$O_2$ (.21) | $CO_2$ (400)<br>$CH_4$ (1) | 16 | Asyn | Modern Earth | 23.5 | .273 |
| (35) 64/1.0 | 1 | 5785 | .984 | $N_2$ (.79)<br>$O_2$ (.21) | $CO_2$ (400)<br>$CH_4$ (1) | 64 | Asyn | Modern Earth | 23.5 | .307 |
| (36) 256/1.0 | 1 | 5785 | .984 | $N_2$ (.79)<br>$O_2$ (.21) | $CO_2$ (400)<br>$CH_4$ (1) | 256 | Asyn | Modern Earth | 23.5 | .328 |
| (37) 365/1.0 | 1 | 5785 | .984 | $N_2$ (.79)<br>$O_2$ (.21) | $CO_2$ (400)<br>$CH_4$ (1) | 365 | 1:1 | Modern Earth | 23.5 | .372 |
| Earth obliquity | (Colose et al., 2018) | | | | | | | | | |
| (38) 1 | 0.79 | 5785 | 1 | $N_2$ | $CO_2$ (1000) | 1 | Asyn | Aqua | 75 | .251 |
| (39) 1 cs | 0.79 | 5785 | 1 | $N_2$ | $CO_2$ (1000) | 1 | Asyn | Aqua | 75 | .450 |
| (40) 2 | 1 | 5785 | 1 | $N_2$ | $CO_2$ (100) | 1 | Asyn | Aqua | 20 | .302 |
| (41) 2 cs | 1 | 5785 | 1 | $N_2$ | $CO_2$ (100) | 1 | Asyn | Aqua | 20 | .483 |
| (42) 3 | 0.79 | 5785 | 1 | $N_2$ | $CO_2(5x10^4)$ | 1 | Asyn | Aqua | 20 | .452 |
| (43) 4 cs | 1 | 5785 | 1 | $N_2$ | None | 1 | Asyn | Aqua | 75 | .427 |
| (44) 5 | 1 | 5785 | 1 | $N_2$ | $CO_2$ (5000) | 1 | Asyn | Aqua | 75 | .312 |
| (45) 6 cs | 0.70 | 5785 | 1 | $N_2$ | $CO_2$ ($10^5$) | 1 | Asyn | Aqua | 75 | .417 |
| (46) 7 | 0.70 | 5785 | 1 | $N_2$ | $CO_2$ (1000) | 1 | Asyn | Aqua | 20 | .500 |
| | | | | | | | | | | |
| Other planets | | | | | | | | | | |
| (47) Early Mars | 0.32 | 5673 | .984 | $CO_2$ (.87) $N_2$ (.12) | $CH_4$ ($10^4$) | 1.02 | Asyn. | Modern Mars | 25.2 | .248 |
| (48) Kepler-1649 b (Kane et al., 2018) | 1.4 | 3200 | 1.013 | $N_2$ | $CO_2$ (376) | 50 | 1:1 | Aqua | 0 | .408 |

[1]$S_{0X}$ = incident stellar flux relative to Earth; $T_{star}$ = stellar effective temperature; $p_{surf}$ = atmospheric surface pressure; $P_{rot}$ = rotation period; $\varphi$ = obliquity; A = Bond albedo; all simulations have either zero or modern Earth eccentricity except for Proxima Centauri b 3:2e30 (e=.30) and Early Mars (e=.093); obliquity simulations labeled "cs" (cold start) were initialized from a snowball Earth; all simulations include $H_2O$ but only trace amounts for Early Mars.

traces of water vapor and ice but is mostly a cold desert planet; and Kepler-1649 b, whose surface temperature stays below 335 K but with stratospheric $H_2O$ mixing ratios an order of magnitude larger than the traditional "moist greenhouse" threshold of Kasting et al. (1993).

## 3. RESULTS

### *3.1 Controls on Bond albedo and equilibrium temperature*

Figure 1 illustrates the major effects on Bond albedo that cause planets to deviate from Earth's $A$ = 0.3. Figures 1a,b and 1c,d are for planets orbiting G and M stars, respectively. Sturtian Earth 4 (Fig. 1a), irradiated by a dimmer Sun than modern Earth and with low $CO_2$ concentration, maintains an equatorial "waterbelt" region of open ocean but is largely covered by sea ice and snow poleward of ~ ±30° latitude, giving it a fairly high Bond albedo ($A$ = .418). In contrast, Archean Earth C (Fig. 1b) has greatly elevated $CO_2$ and $CH_4$ and a very warm, almost ice-free climate despite the faint young Sun, making its albedo darker than modern Earth's ($A$ = .286). Proxima Centauri b Control (Fig. 1c) has a large dayside open ocean area, but this is significantly obscured by optically thick clouds that are typical of synchronously but slowly rotating exoplanets in GCM simulations (Yang et al., 2013). Nonetheless, it has a low Bond albedo ($A$ = .234) because the incident flux from its very cool star is mostly in the near-IR and is strongly absorbed by atmospheric $H_2O$ and $CO_2$, as well as by sea ice, which is darker in the near-IR than the visible (Joshi and Haberle, 2012). Modified Kepler-1649 b (Fig. 1d) is our most highly irradiated M star planet and a good example of the day-night circulation on synchronous slow rotators that produces thick dayside clouds. It thus has a high Bond albedo relative to our other M star cases ($A$ = .408) but much lower than some of our G star planets because of the absorption of the mostly near-IR M star spectrum.

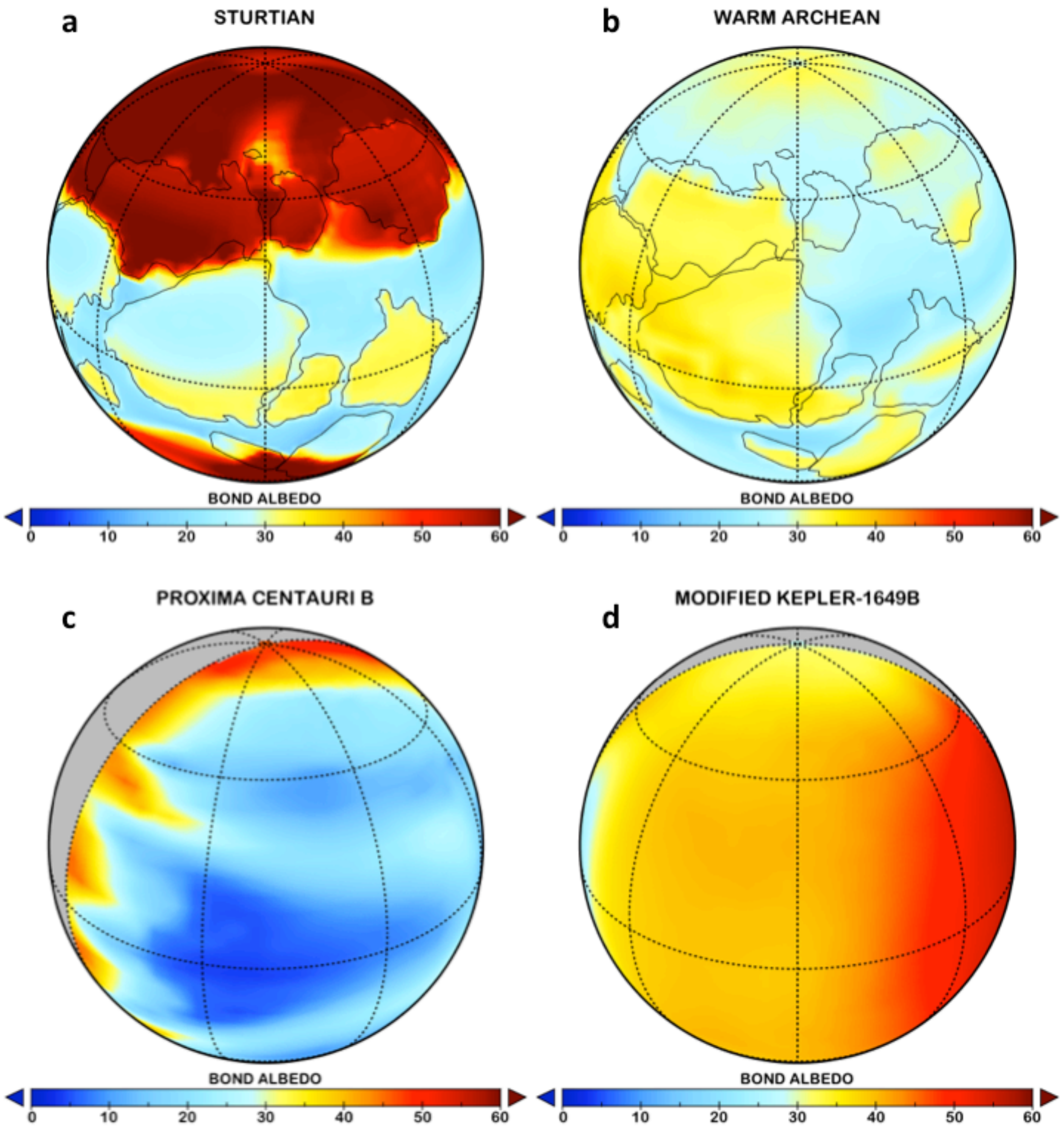

Fig. 1. Bond albedo maps for the (a) Sturtian Earth 4 ($A$ = .418), (b) Archean Earth C ($A$ = .286), (c) Proxima Centauri b Control ($A$ = .234), and (d) modified Kepler-1649 b ($A$ = .408) simulations. Gray areas represent the nightside for synchronously rotating planets.

All of these effects on Bond albedo are evident to varying degrees in the complete ensemble (Fig. 2). $A$ ~ 0.3 is actually quite uncommon in the ensemble, despite it often being the default choice of astronomers (Fig. 2a). For G star planets, the albedo dependence on instellation relative to that received by Earth ($S_{ox}$) exhibits something like a V-shape, with high albedos for planets much more strongly and weakly irradiated than modern Earth and a minimum near the value for modern earth ($S_{ox}$ = 1). This is the result of Earth's surface existing near the triple point of water – at much higher $S_{ox}$ much more water vapor enters the atmosphere and thick clouds develop, obscuring the surface and increasing the Bond albedo, while at much lower $S_{ox}$ sea ice and snow are more widespread, also increasing the Bond albedo (e.g., Fig. 1a).

A few of our G star planets have albedos slightly lower than modern Earth; these are primarily planets with elevated greenhouse gas concentrations and thus less snow/ice (e.g., Fig. 1b), or modestly slow rotators (8 or 16 d period) whose circulation is dominated by a broad Hadley cell rather than day-night contrasts, leading to a narrow equatorial band of reflective clouds rather than the extensive dayside cloud decks on very slowly rotating planets (Yang et al., 2014; Kopparapu et al., 2016; Way et al., 2018). Depending on these other properties, the minimum Bond albedo can occur at $S_{ox}$ values somewhat higher or lower than that of modern

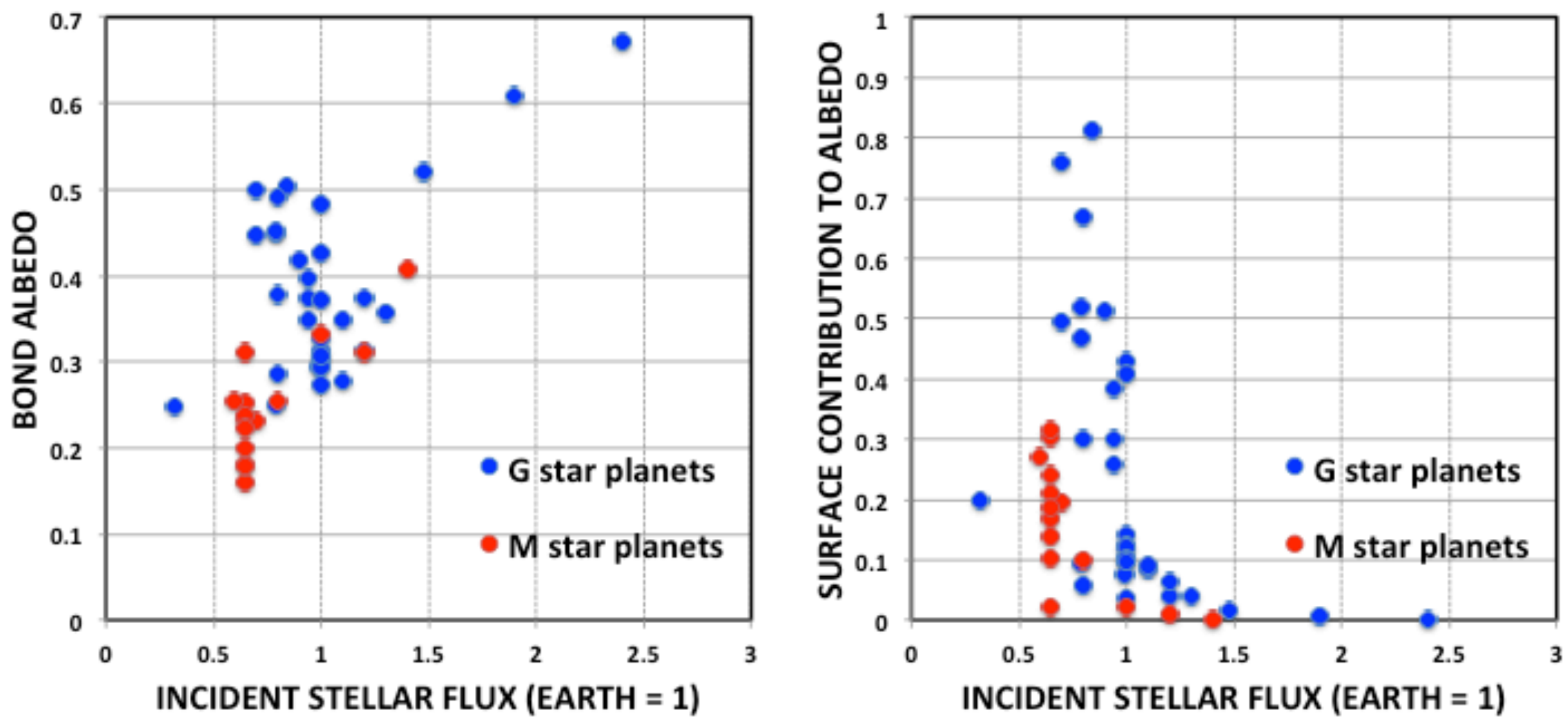

Fig. 2. (a) Bond albedo ($A$) vs. incident stellar flux normalized by Earth's solar constant ($S_{ox}$) for all members of the ensemble. (b) Fractional contribution of the ground albedo to $A$ vs. $S_{ox}$ for the ensemble. Red/blue symbols represent cool/warm star planets.

Earth. Regardless of these other properties, though, it is difficult for a habitable G star planet to have a Bond albedo << 0.3; far from being a representative value, it is close to a lower limit.

Figure 2a also shows that the situation is quite different for M star planets. It is possible for such planets to have albedos higher than Earth if they have high enough instellation and thick dayside clouds (e.g., Fig. 1d), but most of our M star planets have lower albedos than modern Earth. This is due to the mostly near-IR spectrum of M stars and the resulting absorption of incident starlight by atmospheric $H_2O$ and $CO_2$, as well as sea ice. Thus again, $A \sim 0.3$ is not a representative value - it is instead not far from an upper limit for habitable planets orbiting such stars. We emphasize that the behavior described here is only for habitable planets. Planets without atmospheres or surface water can have low Bond albedos, e.g., our dry early Mars case, the darkest G star planet we simulate ($A$ = .248). Likewise, very cold planets may have a hydrohalite crust with a higher near-IR surface albedo (Shields and Carns, 2018) and thus may have slightly higher Bond albedos.

As implied by the discussion above, Bond albedo is the result of contributions from both the atmosphere and surface. Donohoe and Battisti (2011) use upwelling and downwelling solar/stellar flux at the top of the atmosphere and the surface to separate the atmospheric ($A_{atm}$) and surface ($A_{surf}$) contributions to $A$ based on the ground albedo $A_g$ and the fractional reflection $r$ and absorption $a$ of shortwave flux for each pass of the radiation through the atmosphere:

$$A_{atm} = r \tag{1}$$

$$A_{surf} = A_g(1 - r - a)^2/(1 - A_g r) \tag{2}$$

Equation (2) shows that the surface contribution to Bond albedo depends not just on the ground albedo itself, but also on the extent to which starlight reflected from the surface is attenuated by the atmosphere above it. For modern Earth, Donohoe and Battisti (2011) find that $A_{atm} = 0.88A$ and $A_{\mathrm{surf}} = 0.12A$, i.e., despite the fact that Earth's surface is partly visible from space and partly covered by fairly bright desert or very bright sea ice and snow, most of its Bond albedo is due to scattering by clouds and to a lesser extent by Rayleigh scattering of the clear atmosphere. Figure 2b shows $A_{surf}/A$ vs. $S_{ox}$ for the ensemble. All planets with $S_{ox} > 1$ have $A_{surf}/A << 1$, i.e., their Bond albedos are cloud-dominated. Most but not all planets with $S_{ox} < 1$ have $A_{surf}/A > 0.1$. For several G star partial or total snowball planets the surface contribution controls the Bond albedo, but this is not true for any M star planet. The contribution of the surface to Bond albedo for weakly irradiated planets is a function mostly of the opacity of the overlying atmosphere and the albedo of the surface at wavelengths of maximum instellation.

To encapsulate these effects, we "predict" Bond albedo ($A^p$) for the ensemble with two linear regressions of $A$ against $S_{ox}$ and normalized stellar temperature $T'_{star} = T_{star}/4500\ K$, one for strongly irradiated atmosphere-dominated planets and another for weakly irradiated planets with potentially significant surface contributions, guided by the behavior in Figure 2b:

$$A^p = .2623\ S_{ox} + .0494\ T'_{star} + .0008 \qquad (2.40 \geq S_{ox} \geq 0.99) \qquad (3)$$

$$A^p = .1526\ S_{ox} + .2529\ T'_{star} - .0472 \qquad (0.94 \geq S_{ox} \geq 0.32) \qquad (4)$$

The first regression primarily reflects the increasing importance of clouds to albedo as $S_{0x}$ increases. The second regression is more influenced by the effect of stellar temperature on near-IR absorption; it does not capture the effect of increasing albedo on G star planets as $S_{0x}$ decreases due to increasing surface snow and ice because of variations in the effect of the overlying atmosphere due to different greenhouse gas concentrations among the experiments.

Figure 3 shows the predicted albedo vs. actual albedo for each simulation. The RMS error in $A$ is 0.06, vs. 0.12 when $A$ = 0.3 is assumed. We note that the regression coefficients for $T'_{star}$ in equation (3) and for $S_{ox}$ in equation (4) are not statistically significant. In fact single regressions of $A^p$ against $S_{ox}$ for high $S_{ox}$ and against $T'_{star}$ for low $S_{ox}$ give similar RMS errors.

It is evident that $S_{0x}$ and $T_{star}$ have some predictive skill for Bond albedo but that several degeneracies limit this skill. These are visible as clusters of points with the same predicted $A$ but a range of actual $A$. For example, the Proxima Centauri b simulations, all but one having the same $S_{0x}$ and $T_{star}$ and thus the same predicted albedo (0.223), actually range in albedo from 0.161 to 0.310 because of differing greenhouse gas amounts, atmospheric thickness, ocean salinity, spin-orbit state, and presence/absence of exposed land. Two pairs of otherwise identical

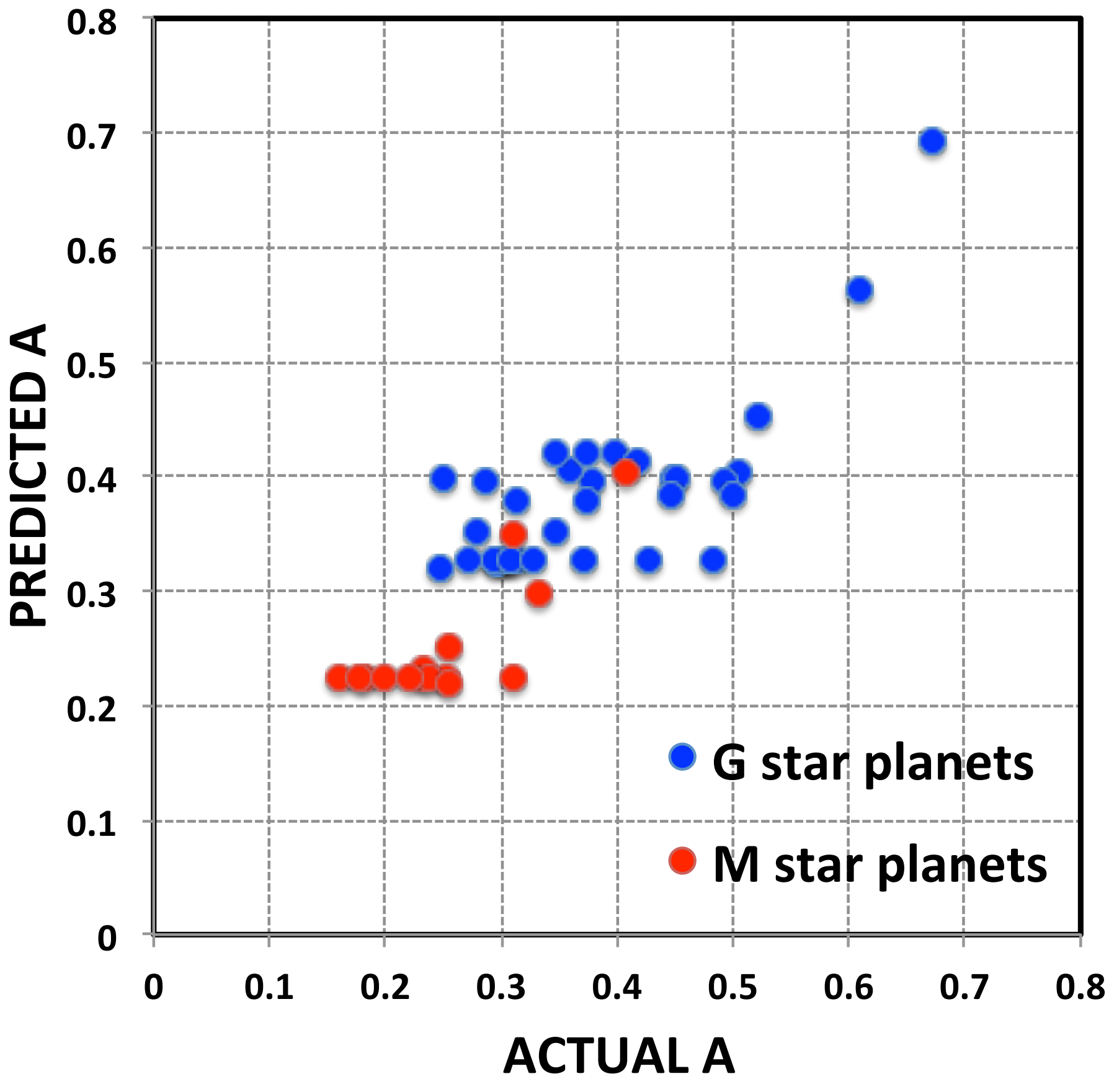

Fig. 3. (a) Predicted ($A^p$) from equations (3,4) vs. actual ($A$) Bond albedo for the ensemble.

Earth simulations that differ only in initial condition (modern Earth vs. snowball state) equilibrate to different climates whose Bond albedos differ by ~0.2. High obliquity planets (75°) tend to have lower albedos than low obliquity planets (20°). The same Earth-like planet with rotation periods from 1-365 d yields Bond albedos ranging from 0.273-0.372. Finally, planets with greater amounts of greenhouse gases tend to be darker.

The resulting predicted equilibrium temperature ($T^p{}_{eq}$) given $A^p$ from equations (3), (4) is

$$T^p{}_{eq} = [S_o(1 - A^p)S_{ox}/4\sigma]^{1/4} \tag{5}$$

where $S_o$ = 1361 W m$^{-2}$ is Earth's solar constant and $\sigma$ the Stefan-Boltzmann constant. The RMS error in $T^p_{eq}$ using $A^p$ is 7 K, vs. 14 K when $A$ = 0.3 is assumed. This improvement is potentially useful, since $S_{ox}$ and $T_{star}$ are known external parameters for every confirmed exoplanet, albeit with non-negligible uncertainties (Brown et al., 2018; Kane, 2018).

*3.2 Inferring surface temperature*

Surface temperature cannot be fully constrained from $S_{0x}$ and $T_{star}$ alone without additional knowledge, e.g., about atmospheric composition and pressure, yet at the moment these are the only two parameters known for any rocky exoplanet that are of first-order importance to habitability. Our ensemble is dominated by Earth-like planets on which $H_2O$ is the dominant absorber but contains examples of un-Earth-like planets to allow us to estimate the errors that might be made in trying to draw inferences about habitability from incomplete information. We consider a surface temperature error of ~ ±20-30°C to be acceptable given the challenge inherent in characterizing exoplanet habitability. Applied to a remote observation of modern Earth as an exoplanet, such an error bar would allow observers to conclude that our planet, if retaining water, at least has open ocean in the tropics and at most is a hot but not runaway greenhouse planet.

Figure 4a shows the relationship between the actual $T_{eq}$ and surface temperature $T_{surf}$ for the GCM ensemble. The two quantities are strongly correlated for this subset of planets, but the standard deviation of $T_{surf}$ over the ensemble is 28°C while it is only 18°C for $T_{eq}$. The biggest outliers tend to be planets with non-condensing greenhouse gas abundances much greater or less than the typical (Earth-like) ensemble member. This illustrates the known inherent limitation of $T_{eq}$ as an indicator of habitability in the absence of direct estimates of $T_{surf}$. A linear regression

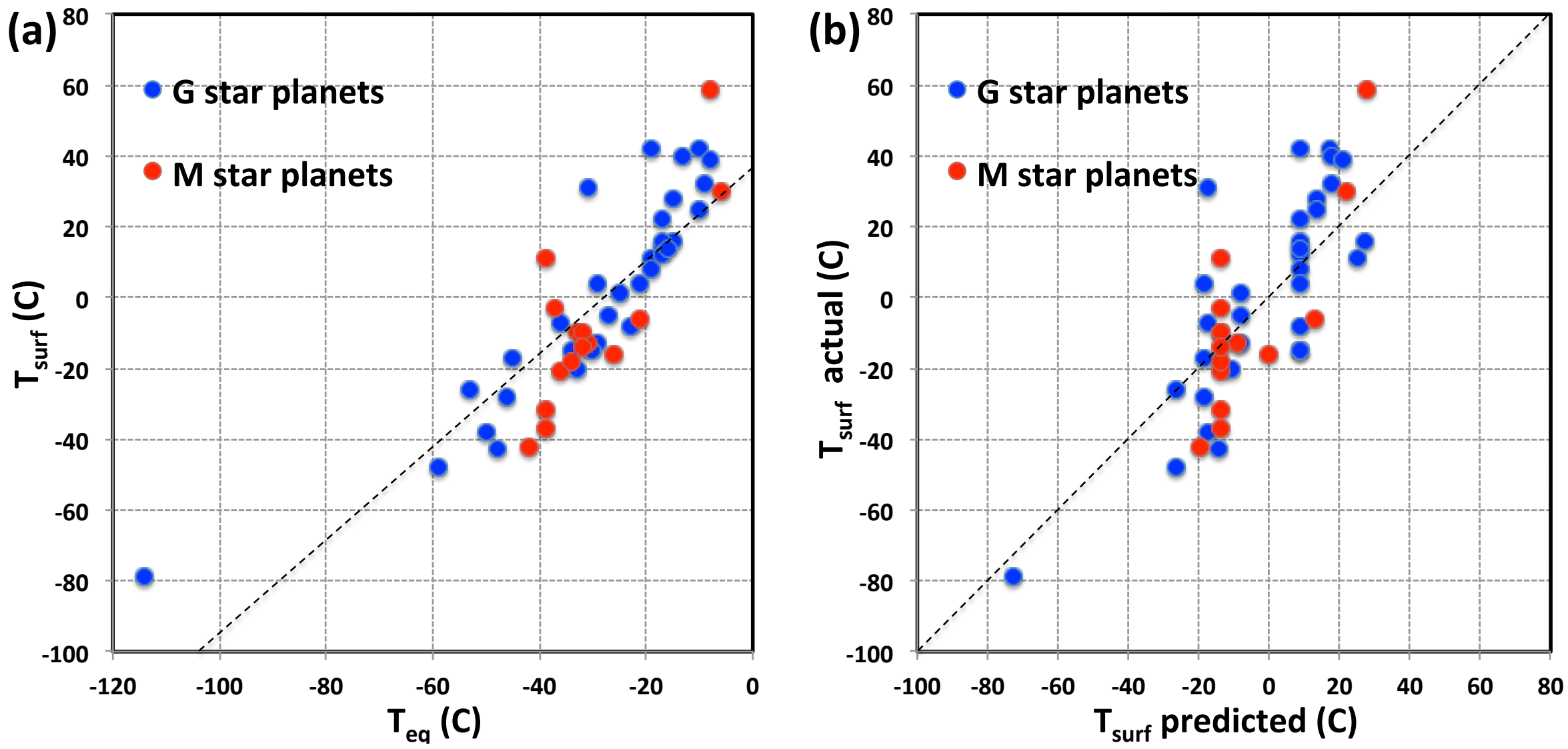

Fig. 4. (a) Actual surface temperature $T_{surf}$ vs. actual $T_{eq}$ (in °C) from the GCM ensemble; the dotted line is equation (6). (b) Actual vs. predicted $T_{surf}$; the dotted line is the 1:1 line.

of the actual $T_{surf}$ vs. actual $T_{eq}$ in Fig. 4a, with both temperatures in °C, gives

$$T_{surf} = 1.327\ T_{eq} + 37.234 = T_{eq} + G_a \tag{6}$$

where $G_a$ is the atmospheric greenhouse effect. Rearranging terms,

$$G_a = 0.327\ T_{eq} + 37.234 = 0.327\ (T_{eq} - T^E_{eq}) + G^E_a\ , \tag{7}$$

where $G^E_a$ = 31.18 °C is the value of $G_a$ for Earth implied by equation (7) for $T^E_{eq}$ = -18.55 °C. This is close to the observed value $G^E_a$ = ~33 °C. Equation (7) indicates that on average, about one-third of the difference in equilibrium temperature between two planets translates into a difference in greenhouse effect. This is most likely due to the negative lapse rate feedback on planets with significant water. On such planets the tropospheric temperature profile is determined to a first approximation by the moist adiabatic lapse rate, which decreases (becomes less steep) with increasing temperature, partly offsetting the positive water vapor feedback (e.g., Soden and Held, 2006). This reduces the error in implied $T_{surf}$ relative to planets without water.

Equations (3), (5), and (6) predict $A$ = 0.327 and $T_{surf}$ = 9.3°C for modern Earth, vs. the observed 0.296 and 15°C, respectively. For modern Mars, a planet outside the outer edge of the habitable zone with a thin mostly $CO_2$ atmosphere and no ocean, equations (4), (5), and (6) yield $T_{surf}$ = -55.9°C, close to its actual -59°C. Applying equation (6) to the ensemble using $A^p$ and $T_{eq}^p$ predicted from equations (3)-(5) gives the result shown in Figure 4b. Equation (6) explains $R^2$ = .65 of the variance in $T_{surf}$ for M star planets and $R^2$ = .63 for G star planets.

Figure 5 shows errors in $T_{surf}$ for the ensemble based on (6) using 3 predictions of $T_{eq}$:

(1) Actual $T_{eq}$ from each simulation, which shows what might be possible when broadband thermal or reflected light phase curves become available for rocky planets.

(2) Predicted $T_{eq}$ from equations (3)-(5), which can be used now with existing information on only $S_{ox}$ and $T_{star}$.

(3) Predicted $T_{eq}$ assuming $A$ = 0.3, i.e., current standard practice for assessing the potential habitability of newly discovered rocky exoplanets.

Overall, the RMS error in $T_{surf}$ is 14°C using the actual $T_{eq}$, 17°C for the $S_{ox}$ - $T_{star}$ predictor, and 21°C for the $A$ = 0.3 assumption, i.e., the regression removes about half the error associated with not knowing the Bond albedo of the planet. Figure 5 shows that if the actual $A$ or $T_{eq}$ are known, $T_{surf}$ is predictable to within 10°C for 27 of the 48 planets and within 20°C for 40 of the 48 planets, if water was retained. At the present time, when only stellar temperature and installation are known, the prediction based on the $S_{ox}$-$T_{star}$ regression is within 20°C of the actual $T_{surf}$ for 34 of 48 planets. This is only slightly better than the default A = 0.3 assumption. The value of a predictor that anticipates tendencies toward high albedos for strongly illuminated planets and low albedos for weakly irradiated planets orbiting cool stars is instead its ability to limit the largest errors: The prediction based on the regression is off by more than 30°C in only 3 cases (about as good as the prediction using the actual $A$), vs. 8 cases when A = 0.3 is assumed.

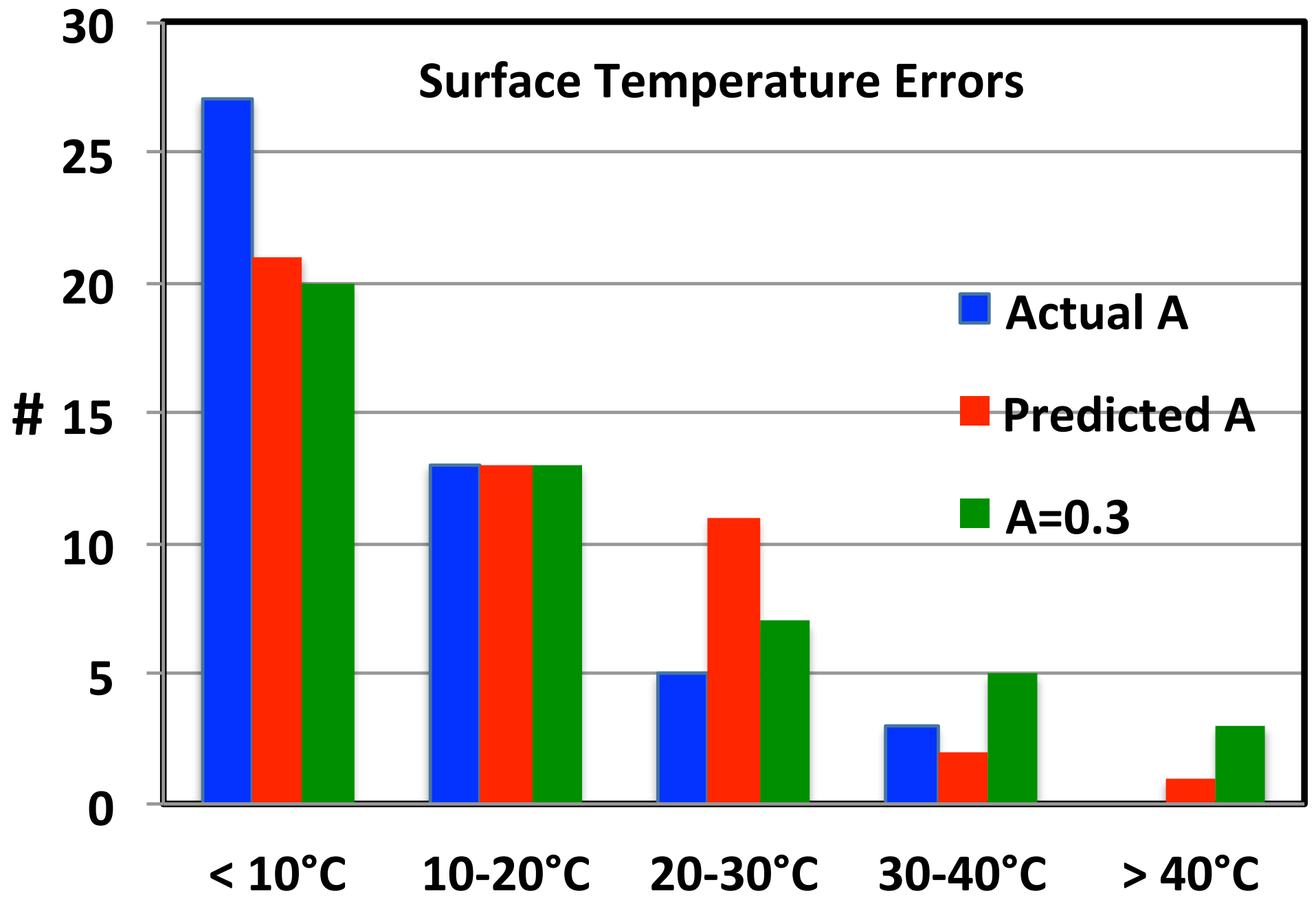

Fig. 5. Frequency histogram of errors in predicted $T_{surf}$ from (6) based on actual $T_{eq}$ (blue), predicted $T_{eq}$ from equations (3)-(5) (red), and predicted $T_{eq}$ assuming $A = 0.3$ (green).

## 4. ASSESSMENTS OF KNOWN EXOPLANETS

To illustrate the potential use of our predictor, we consider some confirmed exoplanets that have been advertised as potentially habitable. For several of these planets, Kane (2018) has recently provided updated estimates of instellation ($S_{ox}$) based on Gaia Data Release 2 (DR2). We exclude Proxima Centauri b, which is part of our GCM ensemble:

1.) *TRAPPIST-1* (Gillon et al., 2017): For TRAPPIST-1 e, equations (4)-(6) predict $T_{surf}$ = -10 or -6°C using the original or DR2 $S_{ox}$ values, respectively, implying a regionally habitable planet, somewhat warmer than our nominal Proxima Centauri b and Sturtian Earth GCM climates. By comparison Wolf's (2018) GCM predicts -32 to +63°C for different compositions; Turbet et al. (2018) reach similar conclusions with a different GCM.

For TRAPPIST-1 f, we predict $T_{surf}$ = -47(-43)°C for the original (DR2) instellation values, respectively, similar to our simulated uninhabitable (at the surface) snowball planets due to its weak $S_{ox}$ = .38(.40), unless several bars of $CO_2$ produce a "maximum greenhouse" habitable planet. This is consistent with GCM estimates (Turbet et al., 2018; Wolf, 2018).

Wolf (2017) finds TRAPPIST-1 d ($S_{ox}$ = 1.14*)* to have a runaway greenhouse, whereas we predict $T_{surf}$ = 21 (23)°C, a warm but habitable planet. This is likely a failure of our simple model. Our regression for $S_{ox}$ > 1 is determined mostly by slowly rotating planets with thick dayside clouds, but TRAPPIST-1 d has a 4 d rotation period if it is synchronous, probably too fast for such clouds to occur, thus destabilizing its climate.

2.) *LHS 1140 b*: Dittmann et al. (2017) classify this as a habitable zone planet, but with $S_{ox}$ =0.46, our predicted $T_{surf}$ = -37°C suggests a near-snowball state. DR2 greatly revised $S_{ox}$ upward to 0.66, though, which increases this planet's prospects for habitability (projected $T_{surf}$ = -13°C), making it potentially similar to Proxima Centauri b and TRAPPIST-1 e.

3.) *Kepler-186 f*: This planet has received attention as a possible cold ($S_{ox}$ = 0.32) but habitable Earth analog. Quintana et al. (2014) suggest that it could sustain liquid water even with an Earth-like atmosphere, while Bolmont et al. (2014) find that 0.5-5 bars of $CO_2$ are required for habitability, using a 1-D model. We predict $T_{surf}$ = -64°C, much colder than our completely glaciated snowball planets. DR2, however, increases $S_{ox}$ to 0.44 and our inferred $T_{surf}$ to -43°C, suggesting marginal habitability if the planet has a thick greenhouse gas atmosphere.

4.) *Kepler-452 b*: This planet receives 11% more incident flux than Earth and was considered a habitable candidate (Jenkins et al., 2015), though its size of 1.6 $R_E$ casts doubt on whether it is rocky. If it is, our predicted $T_{surf}$ = 14°C also suggests habitability.

5.) *Ross-128 b*: Bonfils et al. (2017) characterize this M-star planet as "temperate." They originally estimated $S_{ox}$ = 1.38, but Souto et al. (2018) revise this upward to $S_{ox}$ = 1.79. Our regression predicts $T_{surf}$ = 27 and 33°C for the old and new instellation estimates, respectively, suggesting a planet significantly warmer than Earth, in the class of our warmest early Venus or modified Kepler-1649 b analogs. However, Ross-128 b has a rotation period of only 9.9 days, making it more likely that this planet is hotter than those analogs, at least a moist greenhouse if not a runaway greenhouse planet, if it has an atmosphere and water.

6.) *K2-155 d*: Hirano et al. (2018) use ROCKE-3D to show that this slowly rotating planet, if synchronously rotating with an Earth-like atmosphere but a very small $CO_2$ concentration, would have a moderate climate for $S_{ox}$ as high as 1.5 but does not stabilize at the higher value actually observed ($S_{ox}$ ~ 1.67). For this value, our regression suggests $T_{surf}$ = 31°C, a very warm but not runaway scenario, unlike the ROCKE-3D result. Our closest analog is modified Kepler-1649 b, which is stable for $S_{ox}$ = 1.4 but for which the regression greatly underestimates the actual $T_{surf}$ (see Section 5b).

7.) *GJ 273 b* (Astudillo-Defru et al., 2017): With $S_{ox}$ =1.06, our prediction of $T_{surf}$ = 16°C suggests a potentially Earthlike habitable planet if it is rocky.

8.) *GJ 3293 d* (Astudillo-Defru et al., 2017): $S_{ox}$ = 0.59, and our predicted $T_{surf}$ = -23°C is consistent with a partly habitable planet if it is rocky, analogous to but perhaps somewhat colder than Proxima Centauri b.

9.) *K2-3 d* (Crossfield et al., 2015): $S_{ox}$ = 1.5 and our predicted $T_{surf}$ = 29°C argues for an inner edge, borderline habitable planet like our early Venus 2.4 analog or perhaps the warmest stable K2-155 d simulation of Hirano et al. (2018).

10.) *GJ 625 b*: Suárez Mascareño et al. (2017) describe this as an inner edge planet and we predict $T_{surf}$ = 31°C, but given its red star and high instellation ($S_{ox}$ = 2.1), a runaway greenhouse planet is more likely.

## 5. DISCUSSION

The largest errors in Figure 5 are inherent in the use of the meager information presently available to characterize exoplanets, but the ensemble contains information about other factors that influence Bond albedo and thus equilibrium temperature. Below we estimate the effects of these factors and discuss how some of them might be constrained by future observations.

### *a. Effects of unconstrained parameters on Bond albedo*

Because our ensemble of GCM simulations is sparse and biased toward various clusters of model configurations, we wish to 1) determine non-linearity or linearity in relationships between parameters, and 2) account for confounding interactions. To better understand these, we used the Alternating Conditional Expectations (ACE) algorithm (Breiman and Friedman, 1985). ACE solves for possibly non-linear transformations of variables that maximize their correlation in a generalized additive model (GAM; McCullagh and Nelder, 1989). Details of the approach, its use to identify parameters that have significant influence, and the models that best reduce the deviance from the ensemble albedos, are given in Appendix A. Below we summarize the results.

Although Bond albedo is a non-monotonic function of incident starlight (Fig. 2a), the separate surface and atmospheric contributions (Fig. 2b) are monotonic. Therefore, we performed regressions separately for the two contributions, investigating ways to linearize the relationships to this and other explanatory variables: $T_{star}$ (K), stellar type (G vs. M), orbital period, planet radius, land fraction, obliquity, eccentricity, $CO_2$ and $CH_4$ partial pressure (mb), and interaction terms to account for correlations between parameters. The sum of these two components yields a prediction for Bond albedo that incorporates the effect of each variable on each component. Preliminary ACE fits of the variables were used to identify nonlinearities. To account for some of these we used piecewise fits to the following characteristics: synchronous rotation, low vs. high obliquity, zero vs. non-zero eccentricity, and land fraction. For $CO_2$ and $CH_4$ we regressed on $\log_{10}$ of the amount, which is known to scale linearly with their warming effect (e.g., Hansen et al., 2005). A log transform of $S_{ox}$ also produced improved models.

The final models and the criteria used to define them are detailed in Appendix A. For the atmospheric contribution to $A$ the adjusted $R^2$ is .94, while for the surface contribution it is .74; the resulting predicted Bond albedo has an adjusted $R^2$ of .77. The major explanatory variables for the atmospheric contribution (see Table 3 in Appendix A) in descending order are $S_{ox}$ (which explains 61% of total deviance), planet radius, and stellar type, with minor but statistically significant roles played by $CH_4$ (for M star planets) and orbital period. Several of these probably reflect the sampling of the ensemble rather than any real physical effect. For the surface contribution (see Table 4 in Appendix A), no parameter explains more than 14% of the deviance;

the primary explanatory variables are $S_{ox}$, $CH_4$ (for M star planets), orbital period, and stellar type. The predicted Bond albedo for the GAM model (derived by summing the atmospheric and surface contribution regressions) vs. the actual albedo is shown in Figure 6. Relative to the 2-parameter linear regressions [equations (3) and (4)] in Figure 3, the GAM produces a tighter fit. Figure 7 further illustrates this with histograms of Bond albedo error for the two models.

The GAM approach can also be used to predict $T_{surf}$ as a function of planetary and orbital parameters, instead of the approach based on the physical relationship between $T_{surf}$ and $T_{eq}$ expressed by equations (6) and (7). The details of the model are given in Table 5 in Appendix A. For our ensemble $T_{surf}$ is a strongly linear function of $\log_{10}$ ($S_{ox}$) (56% of total deviance explained), $\log_{10}$ ($CO_2$) (14%), and planet radius (6%), with other contributing parameters totaling 4%, leading to an overall adjusted $R^2$ of .87. Of all these parameters, only $S_{ox}$ and $CO_2$ have an obvious physical connection to $T_{surf}$ – that reflected by equations (6) and (7).

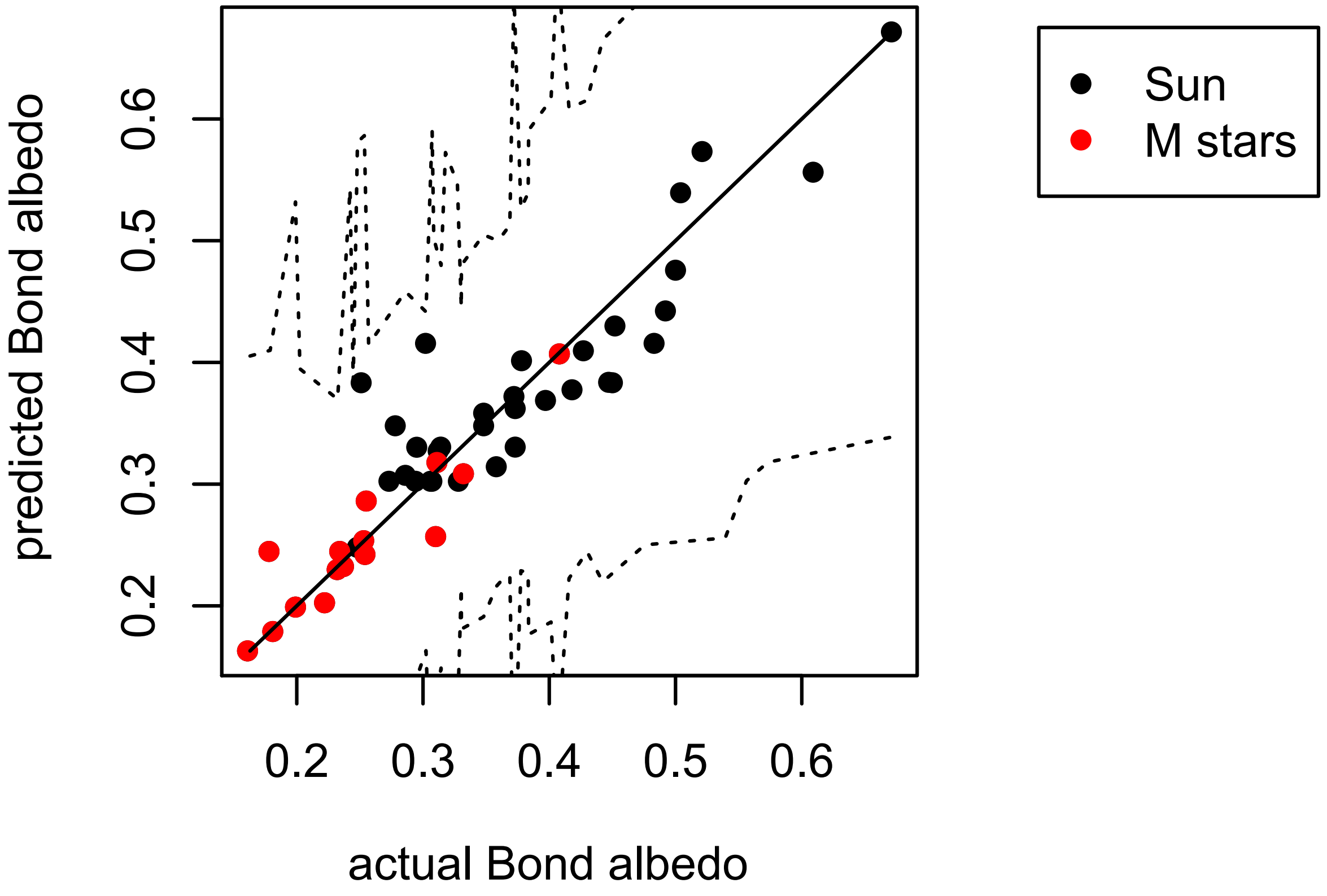

Fig. 6. Predicted vs. actual Bond albedo for the final GAM models. The dashed lines are 95% confidence envelopes.

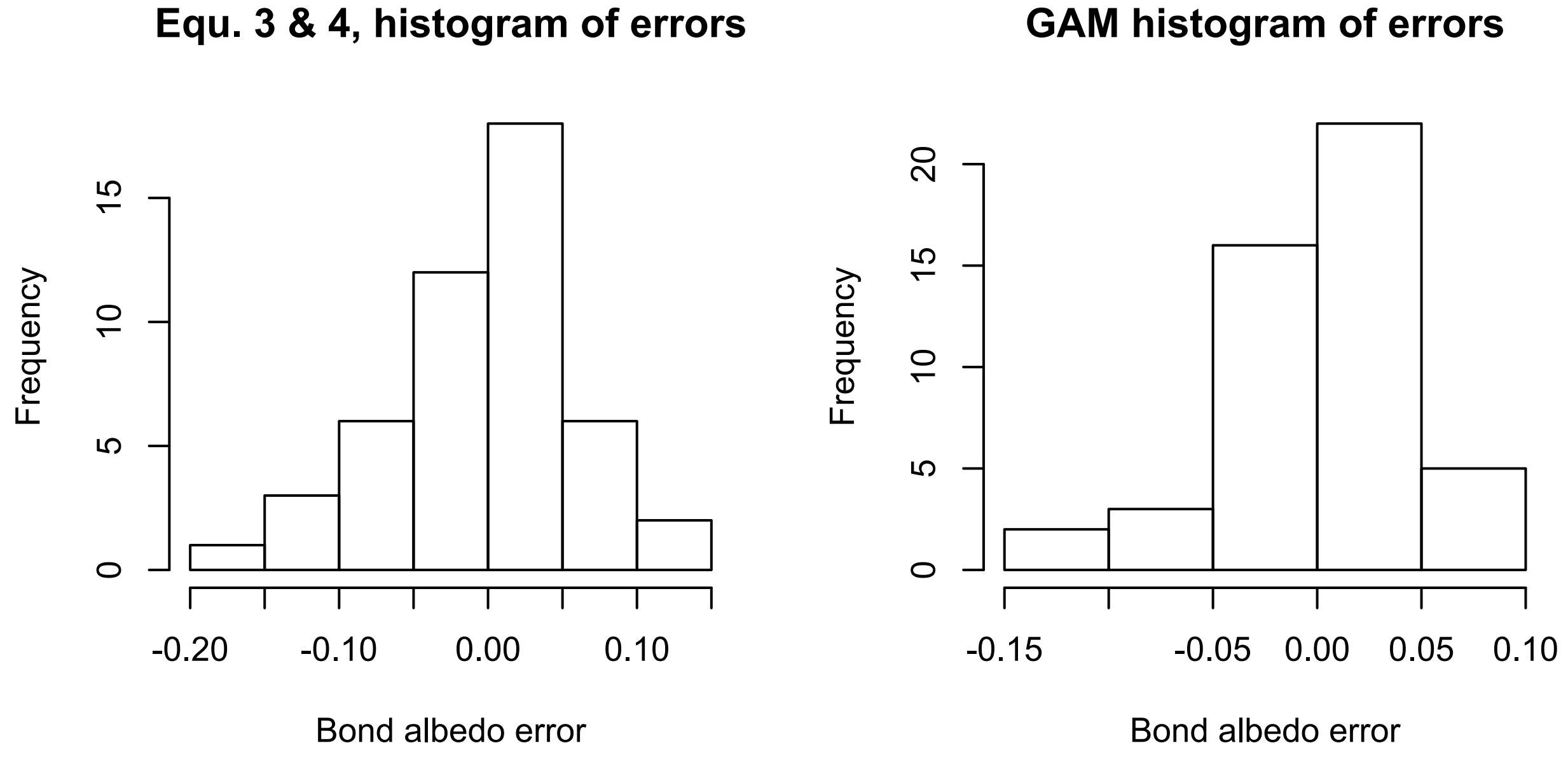

Fig. 7. Frequency histograms of errors in Bond albedo prediction from (left) equations (3) and (4) and (right) the final generalized additive models (GAMs).

*b. Prospects for constraining degeneracies with future observations*

Below we discuss several of the most important degeneracies that affect Bond albedo and surface temperature, and whether observations that might become available in the near term may or may not help narrow the possibilities for specific planets before direct atmospheric composition and surface property characterization of rocky planets becomes possible,.

*i. Greenhouse gas abundances and surface pressure*

Greenhouse gases influence Bond albedo directly, mostly for cool stars, but their biggest effect is the difference they cause between $T_{eq}$ and $T_{surf}$ [equation (6)]. As discussed earlier, some of our largest $T_{surf}$ errors are for planets with high/low concentrations of greenhouse gases, since the difference between $T_{surf}$ and $T_{eq}$ is much larger/smaller for such planets than equations (3)-(6), which are based on mostly Earth-like atmospheres, predict. For synchronously rotating planets, thermal phase curves may help diagnose these errors. The nightside temperature of such planets is highly correlated with their clear-sky greenhouse effect (Yang and Abbot, 2014; Del Genio et al., 2018). The nightside greenhouse effect depends largely on the abundance of non-condensing greenhouse gases, which regulate the condensable gas $H_2O$ (Lacis et al., 2010).

Table 2 shows the $T_{surf}$ errors from equations (3)-(6), the maximum/minimum thermal fluxes to space, and their fractional difference, for 6 planets that differ in atmospheric opacity. They fall into three groups: Optically thin atmospheres (in the thermal IR), atmospheres with primarily $H_2O$ opacity, and optically thick atmospheres with large $CO_2$ opacity. Proxima Thin and GJ 876 0.6 are optically thin for different reasons: The first because its $N_2$ atmosphere is only 100 mb thick and the second because there is only 1 ppmv of $CO_2$ in its weakly illuminated 1 bar atmosphere. Both planets exhibit a large (0.6) fractional flux contrast.

GJ 876 1.2 also has only 1 ppmv $CO_2$ but is illuminated more strongly, leading to a larger dayside $H_2O$ greenhouse effect but less on the cooler (but still warm in an absolute sense) nightside. The Kepler-1649 b case is even more strongly illuminated but with modern Earth concentration of $CO_2$. Both planets maintain a modest nightside clear-sky greenhouse effect and thus a moderate (0.3) contrast between maximum and minimum thermal flux. They are easily distinguished from the optically thin atmospheres by their large nightside thermal emission.

**Table 2**

Day-Night Contrasts and Surface Temperature Errors[1]

| **Simulation** | **$T_{surf}$ error (°C)** | **$F_{max}$ (W m$^{-2}$)** | **$F_{min}$ (W m$^{-2}$)** | **ΔF/F** |
|---|---|---|---|---|
| (8) Proxima Thin | 23 | 260 | 96 | 0.63 |
| (11) GJ 876 0.6 | 22 | 245 | 101 | 0.59 |
| (14) GJ 876 1.2 | -8 | 322 | 212 | 0.34 |
| (48) Kepler-1649 b | -31 | 305 | 214 | 0.30 |
| (9) Proxima Thick | -11 | 197 | 145 | 0.26 |
| (10) Proxima Pure $CO_2$ | -25 | 183 | 158 | 0.14 |

[1]F = thermal emitted broadband flux

The final two planets have considerable $CO_2$ opacity. Proxima Thick is identical to Proxima Thin except that its surface pressure is 10 bars, so its $CO_2$ amount is 100 times larger and the pressure broadening of its absorption lines is much stronger. Proxima Pure $CO_2$ is a 1 bar $CO_2$-only atmosphere. Both planets have a large nightside greenhouse effect, and thus a max-min thermal flux contrast ~2-5 times weaker than our optically thin planets.

With one exception, the error in our predicted $T_{surf}$ goes from strongly positive to increasingly negative as the fractional flux contrast decreases, suggesting that thermal phase curves may provide a useful additional constraint. The exception, Kepler-1649 b, is our hottest and most humid planet, and thus our prediction errs in the same way that it does for the thick $CO_2$ atmospheres while exhibiting more day-night flux contrast.

The largest possible contrast between maximum and minimum emission is expected for synchronously rotating planets that have lost their atmospheres due to various escape mechanisms (Airapetian et al., 2017; Dong et al., 2017; Zahnle and Catling, 2017). It should be possible to differentiate such planets from those with atmospheres using the James Webb Space Telescope (Yang et al., 2013; Kreidberg and Loeb, 2016).

*ii. Photochemical hazes*

One confounding feature of some planets that our ensemble does not represent is a thick photochemical haze that affects Bond albedo independent of the effects of water. Solar System examples are the $H_2SO_4$ solution haze on Venus (Hansen and Hovenier, 1974) and the organic haze on Titan (Hörst, 2017) and perhaps Archean Earth (Zerkle et al., 2012; Arney et al., 2017).

Venus' $H_2SO_4$ haze is probably due to volcanic $SO_2$ emissions (Marcq et al., 2013), as is Earth's episodic volcanic stratospheric haze (Sato et al., 1993). The fact that Earth's volcanic hazes are optically thin and dissipate in a few years while Venus' do not is explained by the presence of liquid water on Earth, which produces storms that wash out stratospheric aerosols soon after they sediment into the troposphere. Loftus (2018) concludes that sulfur chemistry in an oxidized atmosphere with an ocean is probably incompatible with an optically thick haze. Thus we consider it unlikely for a rocky planet within or close to the inner edge of the habitable zone to have a Venus-like Bond albedo unless it is a habitable planet with a thick water cloud.

Titan's organic haze, the product of $CH_4$ photodissociation, is formed at altitudes far above the troposphere and produces a low Bond albedo (~.22) not too different from several habitable G star planets in our ensemble. Indeed Archean Earth, a habitable planet, may have had a somewhat similar haze. Such planets might be identifiable from the spectral dependence of fractal haze particle scattering (Wolf and Toon, 2010) or its unusual phase angle dependence (Garcia Muñoz et al., 2017), but this would leave open the question of whether a habitable surface lay underneath. In principle the same problem might exist for modern Mars-like planets that develop global dust storms, but these should be time-variable and thus distinguishable from hazy or cloudy habitable planets. Furthermore, whether a planet with liquid water and precipitation washout of particulates could ever support a global dust storm is questionable.

*iii. Rotation*

Planet rotation is a key factor in exoplanet habitability but cannot yet be directly observed, although techniques for doing so in the future from photometric variability have been explored (Fujii and Kawahara, 2012; Snellen et al., 2014). Our strongly irradiated habitable planets have high albedos because they rotate slowly (either asynchronous, as in our early Venus and Kepler-1649 b cases, or synchronous rotation, as in our $S_{ox}$ = 1.2 GJ 876 simulation). Rapidly rotating strongly irradiated planets are expected to be darker and enter a runaway greenhouse (e.g., Kopparapu et al., 2016). Habitable planets orbiting cool stars should be tidally locked, so their rotation periods should either match their orbital periods or be in a low order spin-orbit resonance. For weakly irradiated Proxima Centauri b our 3:2 resonance planet is habitable despite its fairly rapid rotation. For a more highly irradiated planet, though, it is possible that resonances higher than 1:1 would compromise habitability. For planets outside the tidal locking radius, there is no way to constrain rotation period without direct observations of it.

*iv. Other parameters*

Bond albedo and thus climate can vary considerably in the face of large obliquity changes, e.g., in Mars' past (e.g., Mischna et al., 2013; Wordsworth et al., 2015; Kite et al., 2017). Furthermore, for G star planets, climate at high obliquity can be bistable (Kilic et al., 2017; Colose et al., 2018), and thus without knowing the dynamical history of such planets (e.g., inward migration across the snow line vs. *in situ* formation), it may be impossible to anticipate habitability. Obliquity is currently unobservable for exoplanets, although it may eventually be inferred when seasonal cycle information on reflected starlight becomes available (e.g., see Kane and Torres, 2017). Planet radius, though a statically significant factor in our GAM, is probably sampled over too small a range by our ensemble (other than the Early Mars simulation) to physically explain any albedo variance. Larger changes in radius can affect climate, though, via the equator-pole temperature gradient (Kaspi and Showman, 2015). Radius is a known quantity for exoplanets detected by the transit method. Eccentricity is not sampled well enough in our ensemble to capture its effect on climate and Bond albedo. It is estimated for some exoplanets but not others, and for sufficiently large values can be degenerate with albedo and obliquity (Barnes et al., 2015; Kane and Torres, 2017,) and it can directly affect albedo by changing the spin-orbit resonance state, as e.g. in our Proxima Centauri b 3:3e30 simulation. The distribution of land and ocean directly influences Bond albedo and indirectly affects it via changes in the climate; retrievals of information for exoplanets are challenging but possible (Fujii et al., 2017b).

## 6. CONCLUSIONS

We have presented a simple approach for synthesizing a fairly large and diverse set of GCM simulations to estimate the Bond albedo, equilibrium temperature, and surface temperature of habitable rocky exoplanets. It implicitly incorporates the effects of dayside cloud shielding on slowly rotating planets, sea ice-snow/albedo feedbacks on weakly irradiated planets, and the enhancement of near-IR absorption of incident starlight by water vapor on planets orbiting cool stars. Earth's existence near the triple point of water makes its albedo close to the minimum that can be expected for a habitable planet orbiting a G star, whereas the low albedo of water vapor and ice in the near-IR gives most habitable M star planets a Bond albedo lower than Earth's.

Relative to the default assumption of a fixed Bond albedo for all exoplanets, our predictor removes about half the error in estimated equilibrium temperature and surface temperature associated with not knowing the Bond albedo and allows surface temperatures to be anticipated to within ~30°C, and usually to much better accuracy, using only knowledge of the stellar flux incident on a planet and the stellar temperature, two known external parameters for every confirmed exoplanet. Given the number of confirmed rocky exoplanets and the much larger population expected to be discovered in the next few years, our technique provides a quick way to identify a small number of the most promising candidates to be targeted for characterization given the reality of long integration times required for useful rocky planet observations and scarce observational resources. At a minimum, we suggest that initial characterizations of the potential habitability of newly discovered rocky exoplanets should not assume a Bond albedo lower than ~0.25 for planets orbiting G stars, or an albedo higher than ~.0.35 for planets orbiting M stars unless they are irradiated more strongly than Earth.

Based on our results, using the Gaia DR2 update, assuming these planets are rocky, and if they retained water, the known exoplanets TRAPPIST-1 e, Kepler-452 b, LHS 1140 b, GJ 273 b, GJ 3293 d, and Proxima Centauri b (which is in our ensemble) have the best chance to be habitable since they can accommodate a fairly large range of greenhouse gas concentrations greater or less than Earth's but still have a moderate climate. The planets TRAPPIST-1 f and Kepler-186 f have a chance to be habitable only if they have a thick enough atmosphere with a greenhouse gas such as $CO_2$ as the major constituent. K2-3 d is a possibly habitable planet at the warm end of the spectrum, while TRAPPIST-1 d, Ross-128 b, K-2 155 d, and GJ 625 b are too highly irradiated and/or too rapidly rotating to have a high likelihood of habitability.

Despite its success, it would make sense to refine the technique in several ways:

1.) We do not include rotation period as a predictor because it is not known for exoplanets. It can be inferred to some extent for tidally locked planets, although such planets may be in a higher order resonance rather than synchronous rotation. Expanding the ensemble to include rapidly rotating (< 10 d), close-in planets that may lose water and develop low albedos would allow us to better anticipate runaway greenhouse conditions.

2.) Any method to infer habitability from $T_{eq}$ alone is limited by the absence of knowledge of the greenhouse effect that determines $T_{surf}$, given current observing capabilities. Our ensemble does not include planets with more than 1 bar of $CO_2$. Greenhouse warming continues to increase up to ~5-8 bars of $CO_2$, beyond which Rayleigh scattering prevents further warming (e.g., Kasting et al., 1993; Wolf, 2018). Whether habitable planets can sustain such thick $CO_2$ atmospheres is not known. It has been suggested that aquaplanets may not have the strong surface weathering sink that removes $CO_2$ (Abbot et al., 2012). Carbon cycle-climate modeling suggests, though, that seafloor weathering may be more effective than previously anticipated (Charnay et al., 2017; Krissansen-Totton et al., 2018), and for thicker ocean "water worlds" seafloor pressure may inhibit $CO_2$ buildup (Kite and Ford, 2018). Early Mars, a planet at the outer edge of the habitable zone, appears to have only had 1-2 bars of $CO_2$ (Kite et al., 2014). Observations that reveal the efficacy of the carbonate-silicate cycle feedback that predicts increasing $CO_2$ retention as illumination decreases (Kasting et al., 1993) would be a useful constraint (e.g., Bean et al., 2017).

3.) Our "ensemble of opportunity" is merely a first step in distilling general information about exoplanet habitability from GCMs. Multiple groups conduct rocky exoplanet GCM simulations (although most do not include dynamic oceans). Together these form a continually growing "grand ensemble" of hypothetical planets that our simple method or the more sophisticated ACE/GAM approach might utilize, both to limit the impacts of inaccuracies inherent to all models, and to exploit the breadth of simulated planets and sampling of parameter space in this ever-expanding storehouse of information that cannot be matched by any single model. Large "perturbed parameter" ensembles of GCM simulations that objectively sample all relevant external parameters, as has been done to estimate uncertainty in projections of terrestrial climate change (e.g., Stainforth et al., 2005), are a logical next step for exoplanet habitability estimates. A repository of such simulations available to the entire community, much like the Coupled Model Intercomparison Project protocol used by Earth GCM modelers for projections of 21st century climate change (Eyring et al., 2016), would accelerate the search for habitable planets.

**ACKNOWLEDGEMENTS.** This research was supported by the NASA Astrobiology Program through collaborations arising from our participation in the Nexus for Exoplanet System Science (NExSS), the NASA Planetary Atmospheres Program, and the GSFC Sellers Exoplanet Environments Collaboration. Computing resources were provided by the NASA High-End Computing (HEC) Program through the NASA Center for Climate Simulation (NCCS) at GSFC.

## Appendix A

## Use of the Alternating Conditional Expectations Algorithm to Develop Generalized Additive Models for Bond Albedo and Surface Temperature

We describe here the procedure for selecting statistical models to incorporate orbital and planetary parameters to predict Bond albedo based the separate contributions of the atmosphere and surface, and also to predict surface temperature. The final models keep only terms that are both significant to 95% confidence in the predicted variable, and whose regression coefficients are estimated to at least 95% confidence. The explanatory variables are:

*Incident stellar flux ($S_{ox}$):* A transform to $log_{10}(S_{ox})$ dramatically improves the AIC score of the models and captures the transition near $S_{ox}$ = 1 between regimes in which the surface contribution to albedo is or is not important.

*Stellar temperature ($T_{star}$)*: In the absence of any K star planets $T_{star}$ is not sufficiently sampled to provide additional information beyond that obtained by separating by stellar type.

*$CO_2$:* A transform to $log_{10}(CO_2 + 10^{-6})$ is based on the well known linear relationship of this quantity to warming. It also alleviates the clustering of points at low concentrations and thus provides better sampling. The $10^{-6}$ offset is sufficiently small to be distinct from actual non-zero values without skewing the distributions while avoiding log(0) instances.

*$CH_4$*: A transform to $log_{10}(CH_4 + 10^{-7})$ is made for similar reasons.

*Obliquity:* Introducing a vector to distinguish obliquity values above and below 20° produces linear relationships to the predicted variables.

*Land fraction:* This parameter is not sufficiently sampled to provide a robust relationship, although there is a discontinuity at Earth's land fraction around which piecewise linear relations can be positive or negative depending on influences from other variables.

*Categorical parameters:* Separating the simulations into subcategories yields distinctly different relations of parameters to the predicted albedo, either parallel but offset, or opposite. We use the following: StellarTypeM to distinguish M vs. G star planets; TLTRUE to distinguish synchronously rotating from other planets; OBL20TRUE to separate high vs. low obliquity planets; E0TRUE to separate planets with zero and non-zero eccentricity; and AQUATRUE to separate ocean-covered planets from planets with exposed land.

*Other parameters:* Planet radius, rotation period, orbital period, and eccentricity are not sufficiently sampled to yield consistent relations but randomly appear as significant in the different fits tested.

The Alternating Conditional Expectations (ACE) algorithm (Breiman and Friedman, 1985), part of the acepack package in R software, solves for possibly non-linear transformations $t_{var}$ of variables that maximize their correlations in a generalized additive model (GAM; McCullagh and Nelder, 1989) of the form

$$t_y(y) = \sum_{i=1}^{n} t_{xi}(x_i) \qquad (8)$$

where $y$ is the predicted variable, $x_i$ are the explanatory variables, and $t_y$ and $t_{xi}$ are the relevant transformations. The transforms are localized and provide no functional parameterizations, but can reveal relationships (positive, negative, non-linearities). They can suggest functions, $g(y) = E(t_y(y))$ or $f_i(x_i) = E(t_{xi}(x_i))$, that can be fit as estimators of the transforms. Those functions can then be used in generalized linear model (GLM) regressions, in which the transformed explanatory variables are treated as linear terms, and the transformation of the response is treated as the link function of the GLM.

A GLM is a simple linear regression used to stratify regressions involving categorical variables to perform a logistical or logit regression that solves for the intercept $b_0$ and coefficients $b_i$ of

$$g_y(y) = \sum_{i=1}^{n} b_0 + b_i f_i(x_i) + \sum_{j=1, i<>j}^{n} b_{i,j} f_i(x_i) f_j(x_j) + \cdots \qquad (9)$$

After examining various ACE experiments we performed a series of GLM regressions, each one removing non-significant parameters or introducing transformations that capture some behavior seen in the ACE transformations, or that normalize the residuals, to improve the fit.

The adjusted coefficient of determination (Adjusted $R^2$) is used as a general measure of goodness of model fit, adjusted for the number of terms. It only increases if a term improves the prediction relative to what would be expected by chance. Starting from

$$R^2 = 1 - \frac{\sigma^2}{\mathrm{Var}(y)} \qquad (10)$$

where $\sigma$ is the residual variability and *Var(y)* the variance of the predictor, the adjusted $R^2$ is

$$\hat{R}^2_{adj} = 1 - \frac{\frac{1}{n-p-1}\sum_{i=1}^{n}(y_i - \hat{y}_i)^2}{\frac{1}{n-1}\sum_{i=1}^{n}(y_i - \bar{y}_i)^2} = \hat{R}^2 - \left(1 - \hat{R}^2\right)\frac{p}{n-p-1} \qquad (11)$$

We also use Student's T-test of standard error estimates as a measure of confidence of the fitted coefficients, and quantify the significance of each parameter in reducing GLM deviance (instead of variance, which assumes normally distributed variables) through the Chi-squared test.

To compare the goodness-of-fit of different GLMs, we use a version of The Akaike Information Criterion (AIC; Akaike, 1981; Sakamoto et al, 1986) to select parsimonious models:

$$\mathrm{AIC} = -2 * log\text{-}likelihood + k * n \qquad (12)$$

where *log-likelihood* is the value for the model, *n* the number of parameters in the model, and *k = 2* is used by AIC to penalize a model for overfitting, the better model having a lower AIC. To quantify whether another model is significantly better, we calculate the likelihood $L_m$ that model *m* minimizes information loss compared to a model with lower AIC ($AIC_{min}$):

$$L_m = \exp\left(\frac{\text{AICmin} - \text{AIC}}{2}\right) \quad (13)$$

We also fit the same GLM as a simple linear regression to estimate $R^2$ and the F statistic, which quantifies the joint significance of the fitted coefficients (where a low probability that the fitted coefficients differ significantly from each other indicates a good model fit, and the statistic can be used to compare different models, similarly to how the AIC score compares GLMs).

The final models used to predict the atmospheric and surface contributions to Bond albedo, and the surface temperature, are given in Tables 3, 4, and 5, respectively. Figure 8 shows the GLM predictions of the surface contribution, atmospheric contribution, and total Bond albedo vs. the actual values simulated by ROCKE-3D, along with histograms of the errors in the prediction of all three albedo components.

**Table 3**

GLM for Atmospheric Contribution to Bond Albedo[1,2]

| **Predictor** | **Estimated Coefficient** | **Significance of Coefficient** | **Deviance Fraction** | **Significance of Variable** |
|---|---|---|---|---|
| $Log_{10}\ S_{ox}$ | 0.368 | *** | 0.614 | *** |
| Planet radius (km) | $6.16 \times 10^{-4}$ | . | 0.174 | *** |
| Stellar type M | -6.94 | . | 0.086 | *** |
| Stellar type M: $\log_{10} CH_4$ | -.0677 | *** | 0.025 | *** |
| Orbital period (d) | 0.0266 | * | 0.018 | *** |
| $Log_{10}$ $CO_2$ (mb) | 0.0112 | ** | 0.012 | ** |
| Obliquity (°):OBL20TRUE | $-6.53x10^{-3}$ | | 0.008 | * |
| $T_{star}$ (K) | $4.67x10^{-3}$ | ** | 0.007 | * |
| TLTRUE | 34.9 | ** | 0.007 | * |
| $T_{star}$ (K):TLTRUE | $-6.02x10^{-3}$ | ** | 0.005 | * |
| Obliquity (°) | 0.0201 | ** | 0.002 | |
| $T_{star}$ (K): orbital period (d) | $-5.83x10^{-6}$ | * | 0.002 | |
| $Log_{10}$ $CH_4$ (mb) | .0613 | *** | 0.001 | |
| Land fraction | -0.994 | . | 0.000 | |
| Eccentricity | 57.2 | ** | 0.000 | |
| AQUATRUE | -0.314 | * | 0.000 | |
| Intercept | -28.4 | *** | - | - |

[1] Significance levels: '***' $\alpha \sim 0$, '**' $\alpha = 0.001$, '*' $\alpha = 0.05$, '.' $\alpha = 0.1$, ' ' $\alpha = 1$

[2] ":" indicates interaction terms between two parameters

**Table 4**

GLM for Surface Contribution to Bond Albedo[1,2]

| Predictor | Estimated Coefficient | Significance of Coefficient | Deviance Fraction | Significance of Variable |
|---|---|---|---|---|
| $Log_{10}$ $S_{ox}$ | -0.569 | *** | 0.144 | *** |
| Stellar type M: $log_{10}$ $CH_4$ | 0.143 | *** | 0.132 | *** |
| Orbital period (d) | -0.0882 | *** | 0.093 | *** |
| Stellar type M | 25.4 | *** | 0.093 | *** |
| $T_{star}$ (K): TLTRUE | 0.0196 | *** | 0.069 | *** |
| $Log_{10}$ $CO_2$ (mb) | -0.0235 | ** | 0.057 | ** |
| Land fraction | 3.7 | *** | 0.057 | ** |
| $T_{star}$ (K):Orbital period (d) | $2.03x10^{-5}$ | *** | 0.053 | ** |
| Orbital per (d): Land fract | $-6.12x10^{-3}$ | ** | 0.037 | ** |
| AQUATRUE | 1.08 | *** | 0.024 | * |
| Obliquity (°) | -0.0615 | *** | 0.018 | . |
| Planet radius (km) | $-2.04 \times 10^{-3}$ | ** | 0.015 | * |
| E0TRUE | -3.44 | * | 0.014 | |
| Obliq (°): OBL20TRUE | 0.0427 | | 0.011 | |
| TLTRUE | -113 | *** | 0.004 | |
| $T_{star}$ (K) | -0.0146 | *** | 0.002 | |
| $Log_{10}$ $CH_4$ (mb) | -0.146 | *** | 0.002 | |
| Eccentricity | -190 | *** | 0.001 | |
| Intercept | 89.7 | *** | - | - |

[1] Significance levels: '***' $\alpha \sim 0$, '**' $\alpha = 0.001$, '*' $\alpha = 0.05$, '.' $\alpha = 0.1$, ' ' $\alpha = 1$

[2] ":" indicates interaction terms between two parameters

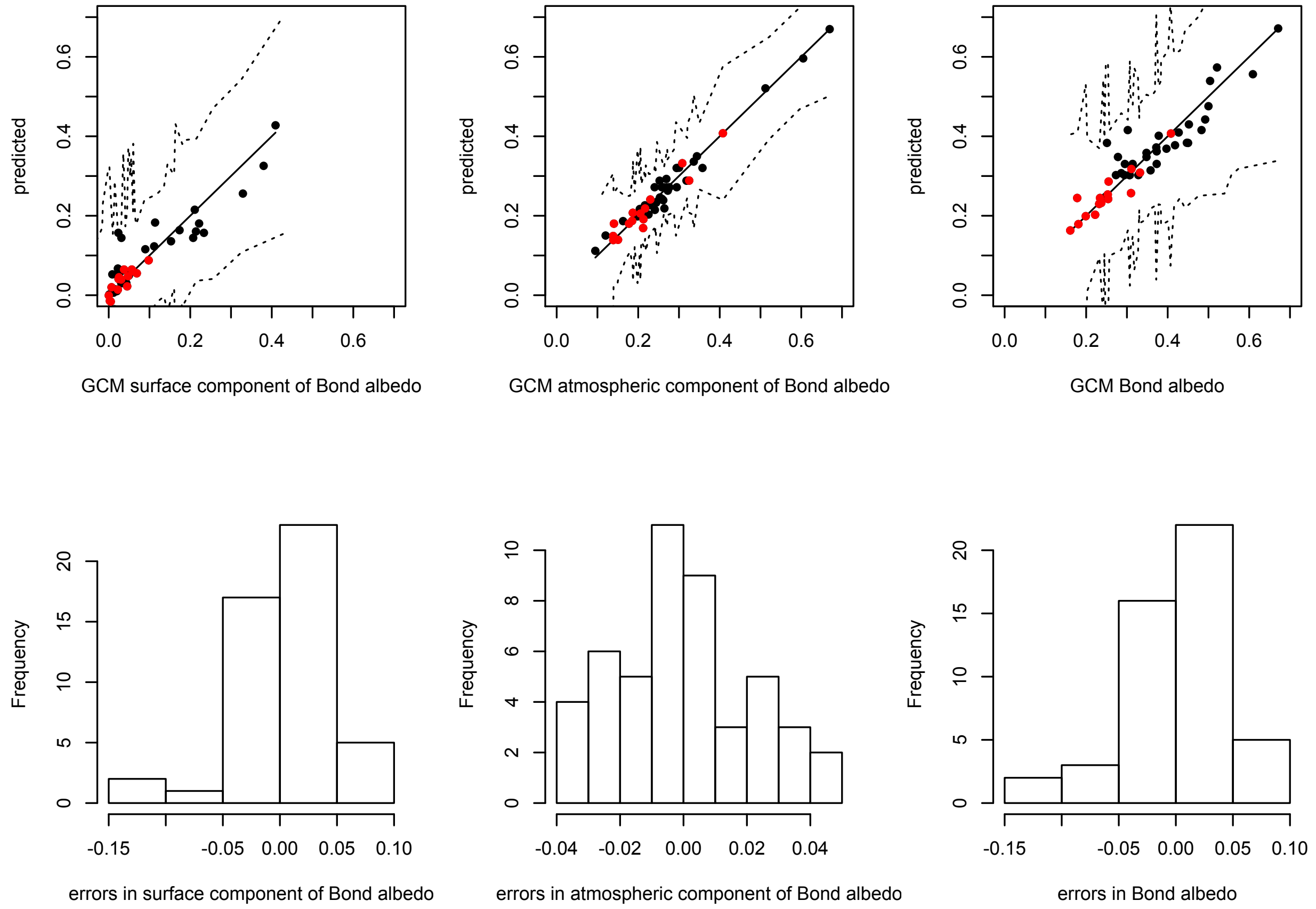

Fig. 8. Upper panels: Predicted vs. actual (left) surface contribution, (center) atmospheric contribution, and (right) total Bond albedo for the GLM. The dashed lines are 95% confidence intervals. Lower panels: Corresponding frequency histograms of errors in the predicted albedos.

**Table 5**

GLM for Surface Temperature[1]

| Predictor | Estimated Coefficient | Significance of Coefficient | Deviance Fraction | Significance of Variable |
|---|---|---|---|---|
| $Log_{10}$ S0X | 159 | *** | 0.557 | *** |
| $Log_{10}$ $CO_2$ (mb) | 8.11 | *** | 0.139 | *** |
| Planet radius (km) | 0.195 | | 0.062 | *** |
| Stellar type M: $log_{10}$ $CH_4$ | -22.4 | *** | 0.042 | *** |
| Orbital period (d) | 14.9 | ** | 0.026 | ** |
| Stellar type M | $-4.43x10^3$ | ** | 0.025 | ** |
| Orbital per (d): Land fract | 1.33 | ** | 0.016 | * |
| E0TRUE | 849 | ** | 0.011 | * |
| $T_{star}$ (K): TLTRUE | -3.03 | *** | 0.011 | . |
| TLTRUE | $1.75x10^4$ | *** | 0.005 | |
| Obliquity (°) | 10.6 | *** | 0.004 | |
| Eccentricity | $3.04x10^4$ | *** | 0.002 | |
| AQUATRUE | -204 | *** | 0.002 | |
| $T_{star}$ (K): Orbital period (d) | $-3.65x10^{-3}$ | ** | 0.002 | |
| $T_{star}$ (K) | 2.27 | *** | 0.001 | |
| Land fraction | -755 | *** | 0.000 | |
| $Log_{10}$ $CH_4$ (mb) | 22.0 | *** | 0.000 | |
| Obliq (°); OBL20TRUE | -12.9 | * | 0.000 | |
| Intercept | $-1.28x10^4$ | *** | - | - |

[1] Significance levels: '***' $\alpha \sim 0$, '**' $\alpha = 0.001$, '*' $\alpha = 0.05$, '.' $\alpha = 0.1$, ' ' $\alpha = 1$

[2] ":" indicates interaction terms between two parameters

**REFERENCES**

Abbot, D. S., Cowan, N. B., and Ciesla, F. J. 2012, *ApJ*, 756, 178

Airapetian, V. S., Glocer, A., Khazanov, G. V., et al. 2017, *ApJL*, 836, L3

Akaike, H. 1981, *JEconometr*, 16, 3

Anglada-Escudé, G., Amado, P. J., Barnes, J., et al. 2016, *Nature,* 536, 437

Astudillo-Defru, N., Foreville, T., Bonfils, X., et al. 2017, *A&A*, 602, A88

Arney, G. N., Meadows, V. S., Domagal-Goldman, S. D., et al. 2017, *ApJ,* 836, 49

Barnes, R., Meadows, V. S., & Evans, N. 2015, *ApJ*, 814, 91

Bean, J. L., Abbot, D. S., & Kempton, E. M.-R. 2017, *ApJL*, 841, L24

Bolmont, E., Raymond, S. N., von Paris, P., et al. 2014, *ApJ*, 793, 3

Bolmont, E., Libert, A.-S., Leconte, J., & Selsis, F. 2016, *A&A*, 591, A106

Bonfils, X., Astudillo-Defru, N., Díaz, R., et al. 2017, *A&A*, 613, A25

Borucki, W. J., Koch, D. G., Batalha, N., et al. 2012, *ApJ*, 745, 120

Boutle, I. A., Mayne, N. J., Drummond, B., et al. 2017, *A&A*, 601, A120

Breiman, L., & Friedman, J. H. 1985, *JAmerStatAssoc*, 80, 580

Brown, A. G. A., Vallenari, A., Prusti, T., et al. 2018, *A&A*, 616, A1

Charnay, B., Forget, F., Wordsworth, R., et al. 2013, *JGR-D*, 118, 1

Charnay, B., Le Hir, G., Fluteau, F., et al. 2017, *EPSL*, 474, 97

Colose, C. M., Del Genio, A. D., and Way, M. J. 2018, in preparation

Crossfield, I. J. M., Petigura, E., Schlieder, J. E., et al. 2015, *ApJ*, 804, 10

Del Genio, A. D., Way, M. J., Amundsen, D. S., et al. 2018, *AsBio*, doi:10.1089/ast.2017.1760

Dittmann, J. A., Irwin, J. M., Charbonneau, D., et al. 2017, *Nature*, 544, 333

Dong, C., Huang, Z., Lingam, M., et al. 2017, *ApJL*, 847, L4

Donohoe, A., & Battisti, D. S. 2011, *JClim*, 24, 4402

Eyring, V., Bony, S., Meehl, G. A., et al. 2016, *GMD*, 9, 1937

Fujii, Y., & Kawahara, H. 2012, *ApJ*, 755, 101

Fujii, Y., Del Genio, A. D., & Amundsen, D. S. 2017a, *ApJ*, 848, 100

Fujii, Y., Lustig-Yeager, J., & Cowan, N. B. 2017b, *AJ*, 154, 189

Garcia Muñoz, A., Lavvas, P., & West, R. A. 2017, *Nature Astron*, 1, 0114

Gillon, M., Triaud, A. H. M. J., Demory, B.-O., et al. 2017, *Nature*, 542, 456

Haqq-Misra, J., Wolf, E. T., Joshi, M., Zhang, X., & Kopparapu, R. K. 2018, *ApJ*, 852, 67

Hansen, J. E., & Hovenier, J. W. 1974, *JAtmosSci*, 31, 1137

Hansen, J., Sato, M., Ruedy, R., et al. 2005, *JGR-D*, 110, D18104, doi:10.1029/2005JD005776

Hirano, T., Dai, F., Livingston, J. H., et al. 2018, *AJ*, 155, 124

Hörst, S. M. 2017, *JGR-E*, 122, 432

Jansen, T., Scharf, C., Way, M., and Del Genio, A. 2018, submitted to ApJ, arXiv:1810.05139

Jenkins, J. M., Twicken, J. D., Batalha, N. M., et al. 2015, *AJ*, 150, 56

Joshi, M. M., and Haberle, R. M., 2012, *Asbio*, 12, 3

Kane, S. R., Hill, M. L., Kasting, J. F., et al. 2016, *ApJ*, 830, 1

Kane, S. R., & Torres, S. M. 2017, *AJ*, 154, 204

Kane, S. R., 2018, *ApJL*, 861, L21

Kane, S. R., Ceja, A. Y., Way, M. J., & Quintana, E. V. 2018, *ApJ*, accepted, arXiv:1810.10072

Kaspi, Y., & Showman, A, P. 2015, *ApJ*, 804, 60

Kasting, J. F., Whitmire, D. P., & Reynolds, R. T. 1993, *Icarus*, 101, 108

Kilic, C., Raible, C. C., & Stocker, T. F. 2017, *ApJ*, 844, 147

Kite, E. S., Williams, J.-P., Lucas, A., & Aharonson, O. 2014, *Nature Geosci.*, 7, 335

Kite, E. S., Sneed, J., Mayer, D. P., & Wilson, S. A. 2017, *GRL*, 44, doi:10.1002/2017GL072660

Kite, E. S., & Ford, E. B. 2018, *ApJ*, 864, 75

Kopparapu, R. k., Wolf, E. T., Haqq-Misra, J., et al. 2016, *ApJ*, 819, 84

Kopparapu, R. k., Wolf, E. T., Arney, G., et al. 2017, *ApJ*, 845, 5

Kreidberg, L., and Loeb, A. 2016, *ApJL,* 832, L12

Krissansen-Totton, J., Arney, G. N., & Catling, D. C. 2018, *PNAS*, 115, 4105-4110

Lacis A. A., Schmidt G. A., Rind, D., & Ruedy R. A. 2010, *Science*, 330, 356

Leconte, J., Forget, F., Charnay, B., Wordsworth, R., & Pottier, A. 2013, *Nature*, 504, 268

Lewis, N. T., Lambert, F. H., Boutle, I. A., et al. 2018, *ApJ,* 854, 171

Loftus, K. 2018, Abstract, *Exoplanets II*, 2-6 July 2018, Cambridge, UK

Marcq, E., Bertaux, J.-L., Montmessin, F., and Belyaev, D. 2013 *Nature Geosci.*, 6, 25

McCullagh, P., & Nelder, J. A. 1989, *Generalized Linear Models*, Chapman and Hall, London, 532 pp.

Mischna, M.A., Baker, V., Milliken, R., et al. 2013, *JGR-E*, 118, 560

Quintana, E. V., Barclay, T., Raymond, S. N., et al. 2014, *Science*, 344, 277

Sakamoto, Y., Ishiguro, M., & Kitagawa, G. 1986, *Akaike Information Criterion Statistics*, D. Reidel Publishing Co., Springer Netherlands, 290 pp.

Sato, M., Hansen, J. E., McCormick, M. P., & Pollack, J. B. 1993, *JGR-A*, 98, 22987

Shields, A. L., Meadows, V. S., Bitz, C. M., et al., 2013, *AsBio*, 13, 715

Shields, A. L., Bitz, C. M., Meadows, V. S., Joshi, M. M., & Robinson, T. D. 2014, *ApJL,* 785, L9

Shields, A. L., & Carns, R. C. 2018, *ApJ*, 867, 11

Snellen, I., Brandl, B., de Kok, R., et al. 2014, *Nature*, 509, 63

Soden, B. J., & Held, I. M. 2006, *JClim*, 19, 3354

Sohl, L. E., and Chandler, M. A., 2007, *Deep-Time Perspectives on Climate Change: Marrying the Signal from Computer Models and Biological Proxies*, Micropalaeontological Society Special Publication #2, 61

Souto, D., Unterborn, C. T., Smith, V. V., et al. 2018, *ApJL*, 860, L15

Stainforth, D. A., Aina, T., Christensen, C., et al. 2005, *Nature*, 433, 403

Suárez Mascareño, A., González Hernández, J. I., Rebolo, R., et al. 2017, *A&A*, 605, A92

Turbet, M., Leconte, J., Selsis, F., et al. 2016, *A&A*, 596, A112

Turbet, M., Bolmont, E., Leconte, J., et al. 2018, *A&A*, 612, A86

Way, M. J., Del Genio, A. D., Kiang, N. Y., et al. 2016, *GRL*, 43, 8376

Way, M. J., Aleinov, I., Amundsen, D. S., et al. 2017, *ApJS*, 231, 12

Way, M. J., Del Genio, A. D., Kelley, M., et al. 2018, *ApJS*, 239, 24

Wolf, E. T., & Toon, O. B. 2010, *Science*, 328, 1266

Wolf, E. T., & Toon, O. B. 2014, *GRL*, 41, doi:10.1002/2013GL058376

Wolf, E. T., & Toon, O. B. 2015, *JGR-D*, 120, 5775

Wolf, E. T., 2017, *ApJL*, 839, L1

Wolf, E. T. 2018, *ApJL*, 855, L14

Wolf, E. T., Shields, A. L., Kopparapu, R. K., et al. 2017, *ApJ*, 837, 107

Wordsworth, R. D., Kerber, L., Pierrehumbert, R. T., et al. 2015, *JGR-E*, 120, 1201

Yang, J., Cowan, N. B., & Abbot, D. S. 2013, *ApJL*, 771, L45

Yang, J., & Abbot, D. S. 2014, *ApJ*, 784, 155

Yang, J., Boué, G., Fabrycky, D. C., & Abbot, D. S. 2014, *ApJL*, 787, L2

Zahnle, K. J., & Catling, D. C. 2017, *ApJ*, 843, 122

Zerkle, A., Claire, M. W., Domagal-Goldman, S. D., et al. 2012, *Nature Geosci.*, 5, 359